\documentclass[aip,reprint]{revtex4-1}

\usepackage{graphicx} 
\usepackage{epsfig}
\usepackage{amsmath} 
\usepackage{amsthm} 
\usepackage{amssymb}	
\usepackage{physics}
\usepackage{hyperref} 
\hypersetup{
    colorlinks,
    citecolor=blue,
    filecolor=black,
    linkcolor=red,
    urlcolor=blue
}

\newcommand{\dmin}{D^2_{\min}}
\newcommand{\delmin}{\Delta^2_{\min}}
\DeclareMathOperator*{\CDF}{\expval{\text{CDF}(S_{i_{\max}})}}
\DeclareMathOperator*{\argmin}{arg\,min}

\usepackage[english]{babel}
\usepackage{blindtext}
\usepackage{makecell}
\usepackage{tikz}

\begin{document}


\title{Learning-based approach to plasticity in athermal sheared amorphous packings: Improving softness
} 




\author{Jason W. Rocks}
\author{Sean A. Ridout}
\author{Andrea J. Liu}
\affiliation{Department of Physics and Astronomy, University of Pennsylvania, Philadelphia, PA 19104, USA}



\date{\today}

\begin{abstract}
The plasticity of amorphous solids undergoing shear is characterized by quasi-localized rearrangements of particles. 
While many models of plasticity exist,
the precise relationship between plastic dynamics and the structure of a particle's local environment remains an open question. 
Previously, machine learning was used to identify a structural predictor of rearrangements, called ``softness.''
Although softness has been shown to predict which particles will rearrange with high accuracy,
the method can be difficult to implement in experiments where data is limited and the combinations of descriptors it identifies are often difficult to interpret physically.
Here we address both of these weaknesses, presenting two major improvements to the standard softness method.
First, we present a natural representation of each particle's observed mobility, 
allowing for the use of statistical models which are both simpler and provide greater accuracy in limited data sets.
Second, we employ persistent homology as a systematic means of identifying simple, topologically-informed, structural quantities that are easy to interpret and measure experimentally. 
We test our methods on two-dimensional athermal packings of soft spheres under quasi-static shear.
We find that the same structural information which predicts small variations in the response is also predictive of where plastic events will localize. 
We also find that excellent accuracy is achieved in athermal sheared packings using simply a particle's species and the number of nearest neighbor contacts.
\end{abstract}


\maketitle 

\section{Introduction}

Machine learning has proven effective in identifying a structural quantity, softness, which predicts plastic rearrangements in solids~\cite{Cubuk2015,Schoenholz2016,Sharp2018}.
In crystals, defects in crystalline order such as dislocations and grain boundaries -- around which rearrangements are known to localize~\cite{Taylor1934, Taylor1934a}  -- are typically characterized by high softness~\cite{Sharp2018}.
In disordered solids, softness succeeds remarkably well at addressing the longstanding challenge of identifying a structural indicator of a particle's propensity to rearrange.
In particular, in supercooled liquids, the probability that a particle will rearrange depends approximately exponentially on the particle's softness, 
spanning several orders of magnitude~\cite{Schoenholz2016}.
Although softness is highly predictive in a wide range of systems studied in both simulations and experiments~\cite{Cubuk2015, Schoenholz2016, Cubuk2017}, however,
it still suffers from some significant drawbacks.

The first drawback is a practical one. Calculation of softness is significantly constrained by the need for training examples of rearranging particles, which constitute a very small fraction of the total number of particles in the system~\cite{Schoenholz2016}.
This has the effect of substantially increasing the number of independent configurations needed,
reducing the method's practical use for analyzing limited experimental data.

The second drawback is a scientific one. Although softness yields insight into the underlying physics of glassy systems and can be considered a quantification of the old idea of a cage~\cite{Schoenholz2016}, 
its meaning in terms of the local structure can be difficult to interpret because it is defined in terms of a large number of local parameters.
This diminishes the insight that it can provide for the development of a structural theory of plasticity from first principles.
 
To calculate softness, a support vector machine (SVM) is trained to sort particles into one of two classes based on their current local structure: 
particles that are likely to participate in a rearrangement in the future (those with ``high softness'' environments) and particles that are not likely to participate in a rearrangement (those with ``highly negative softness'' environments). 
To train the classifier, examples of non-rearranging and rearranging particles (which we will refer to as ``rearrangers'' and ``non-rearranger'', respectively) must first be identified by observing a series of configurations undergoing rearrangement events in either simulation or experiment.
The SVM then attempts to find a hyperplane which best separates the two classes of particles within a high-dimensional space of structural descriptors.
These descriptors are derived from each particle's local pair correlation function.
Softness is computed as the signed distance from the hyperplane, 
with particles located far from the hyperplane in the positive direction being softer, and therefore more likely to rearrange,
while particles located far from the hyperplane in the negative direction considered to be harder and less likely to rearrange.

Here we propose two refinements to the softness calculation, addressing both drawbacks of the previous approach.
First, we consider local variations in the displacement far from a plastic event to define a dynamical quantity which effectively characterizes a particle's susceptibility to rearrangements, or mobility.
Since we can calculate this quantity for each particle in a system, we avoid the problem of having to choose examples of relatively mobile and immobile particles,
converting the softness problem into one of regression rather than classification.
This has the effect of greatly increasing the amount of data available from a single configuration and thereby improving performance when applying this technique to experimental systems or simulations with limited data.

Second, we use persistent homology, a form of topological data analysis, 
to systematically define a set of simple local structural parameters in a physically meaningful way, 
eliminating much of the guesswork. 
We demonstrate how to combine persistent homology with a machine-learning-based approach to identify a new version of softness that captures correlations between the dynamics and the local topological structure for each particle.
We compare both aspects of our new approach with current methods to compute softness and demonstrate that it is just as effective.
We find that the same structural information which predicts local fluctuations in the displacement field is also predictive of rearrangements.
Furthermore, we find that excellent accuracy can be achieved with very few structural descriptors, in this case simply a particle's species and the number of nearest neighbors contacts.

This article is organized as follows:
In Sec.~II, we describe our process for generating configurations of particles at the onset of rearrangement.
In Sec.~III, we describe how particle dynamics are characterized to define the original softness, and introduce our new measure of effective particle mobility.
We explain how the choice of characterization determines which type of statistical model is appropriate, which in turn affects the accuracy of the softness method when data is limited.
In Sec.~IV, we apply persistent homology to our configurations and interpret the results of the procedure.
Next, Sec.~V shows how to find correlations between the dynamical and structural characterizations we have developed and uses the resulting insight to define a new set of structural descriptors.
Finally, Sec~VI describes the results of our analysis with further discussion in Sec.~VII.

\section{Onset of Plastic Rearrangements}

To analyze the relationship between structure and dynamics which underlies plasticity, 
we generate an ensemble of configurations of particles undergoing plastic rearrangement.
We first create a collection of jammed packings of soft spheres and then athermally shear each configuration until the onset of rearrangement.

Prior to shearing, we prepare each configuration using standard methods from the study of jamming~\cite{OHern2003}.
We start by placing $N = 2^{14}$ particles at random positions in a square periodic box in $d=2$ dimensions. 
To reduce the probability of creating packings with crystalline structures, 
we choose a $50\%:50\%$ bidisperse mixture of particles with radii  of $\sigma$ and $1.4\sigma$, 
where $\sigma$ is the radius of the smaller particles size.
The size of the box is chosen to achieve an average packing fraction of $\phi = 0.95$, 
well above the jamming density for this ratio of radii.
Particles interact according to a finite-ranged soft pairwise (Hertzian) potential defined by
\begin{align}
V(r_{ij}) &= \left\{
\begin{array}{ll}
\frac{2}{5} \epsilon\qty(1 - \frac{r_{ij}}{R_i+R_j} )^\frac{5}{2} & r_{ij} < R_i + R_j\\
0 & r_{ij} \geq R_i + R_j
\end{array}
\right.\label{eq:potential}
\end{align}
where $R_i$ is the radius of particle $i$,  $r_{ij}$ is the distance between particles $i$ and $j$, and $\epsilon$ sets the energy scale.
Next, we apply the FIRE algorithm to find a local minimum of the energy for the configuration, i.e. a mechanically stable state~\cite{Bitzek2006}. 
We generated approximately $400$ independent configurations in this way, each from a different set of random initial conditions.

\begin{figure}[t!]
    \centering
    \includegraphics[width=1.0\linewidth]{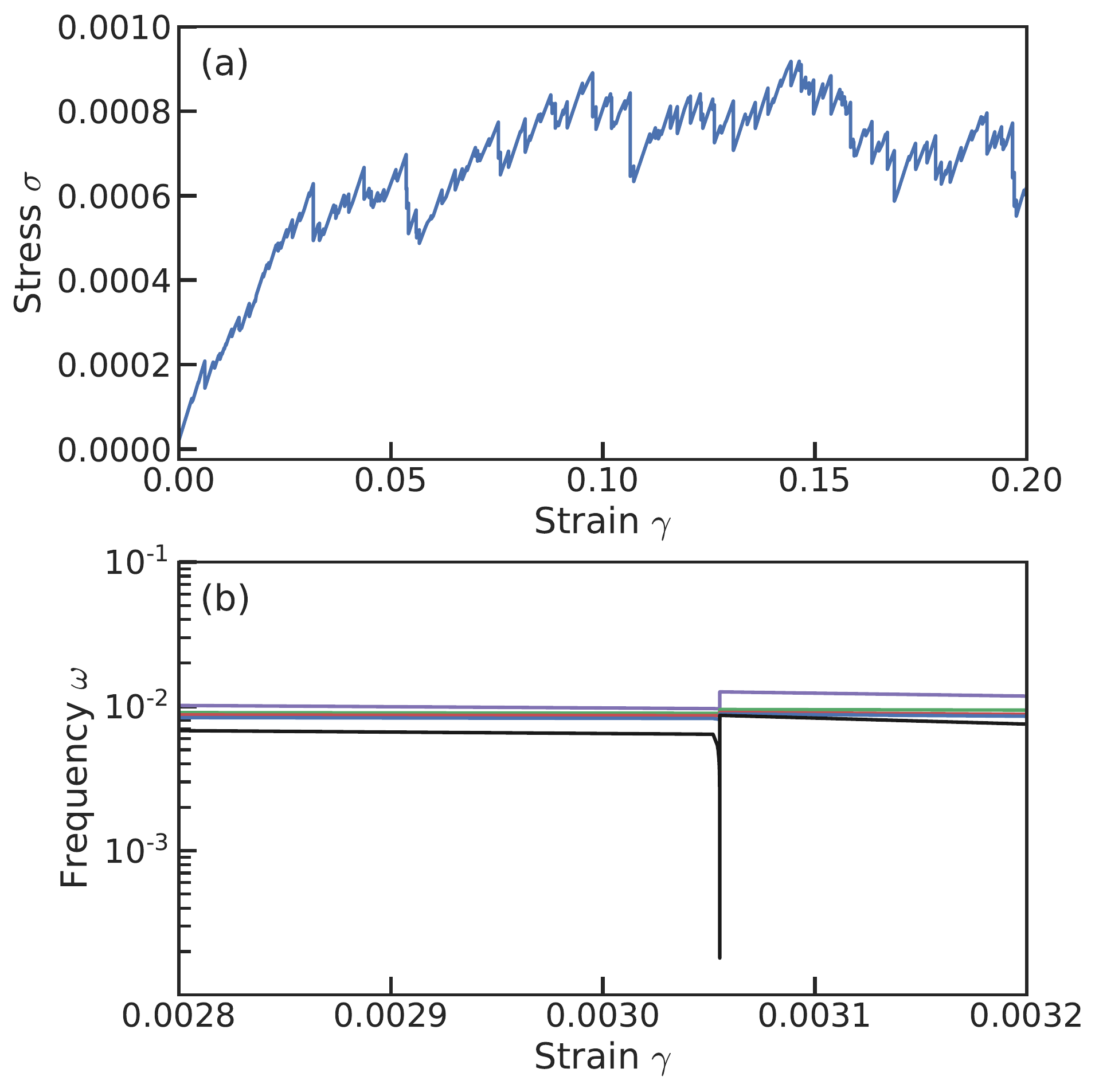}
    \caption{
(a) A typical stress-strain curve for a packing in our simulations. Elastic branches are broken up by sudden plastic events where the shear stress drops; it is the initial rearrangement during these events which we study.
(b)  The lowest 5 normal mode frequencies of the dynamical matrix near one of these plastic events, showing that a single mode frequency goes to zero. The set of displacements described by this critical mode is the initial rearrangement for this plastic event.
    }
    \label{fig:stress}
\end{figure}

Next, we examine each of these configurations under athermal quasistatic shear,
broken up into a sequence of small strain steps.
At each step, we apply a simple shear strain of $10^{-5}$ by changing the shape of the simulation cell and re-minimize the energy to find a new stable state.
Before energy minimization takes place, we make an educated guess as to where the particles will move up to linear order.
During the first step, particles are displaced affinely according to the applied global strain,
while in subsequent steps, the particles are instead moved along the non-affine displacement field produced by the previous strain step.
Since the elastic response of this model is almost piece-wise linear, 
this produces a state closer to the new energy minimum and reduces the time spent on minimization. 

As the system is strained, the shear stress and energy rise; after a sufficient amount of shear strain, the two drop suddenly, as shown in Fig.~\ref{fig:stress}.
These stress drops correspond to plastic (irreversible) events in which particles often change neighbors~\cite{Maloney2006}.
When such an event is detected, we back up to the configuration prior to the event, and approach it a second time with a smaller strain step size. 
This process is repeated until the event is passed through with a strain step size of $10^{-12}$. 
At this very small strain step, the main source of numerical error becomes the finite error of the minimization algorithm, rather than the finite strain step.
We note that similar algorithms have been used in previous work~\cite{VanDeen2014, Morse2020}.

Our goal is to identify which particles are structurally predisposed to rearrange during one of these plastic events. 
However, these events can often involve a sequence of many smaller rearrangements as the movements of some particles can induce the movements of others, 
resulting in an avalanche of rearrangements~\cite{Maloney2006}.  
A particle with a relatively hard structure may become softer during this sequence of rearrangements, 
thus the structure at the beginning of the event should not be expected to strongly predict motion towards the end of the event.

Therefore, we specifically try to identify structures which correlate with the initial motion at the beginning of an event, following earlier work~\cite{Manning2011}.
At the onset of a plastic event, the system becomes linearly unstable toward a single direction in the $dN$ dimensional space of particle coordinates, denoted $\vec{x}_i$ for particle $i$; 
this direction corresponds to the onset of the initial rearrangement~\cite{Maloney2006}. 
We identify this direction by diagonalizing the Hessian (i.e. the matrix of second derivatives of the energy or dynamical matrix)
and identify the eigenvector whose eigenvalue goes to zero at the onset of instability, as shown in Fig.~\ref{fig:stress}.  
We denote this ``critical mode'' $\vec{u}$, with the displacement of particle $i$ denoted $\vec{u}_i$.  
An example of such a critical mode is shown in Fig.~\ref{fig:rearrangement}.
During each shear trajectory, we record the first 10 particle rearrangement events.
We utilize the first event in each trajectory for our analyses and the remaining events to identify examples of particles that are unlikely to rearrange in the future for the classification-based approach (see next section).

\section{Dynamical Characterization and Supervised Learning Strategy}

The softness method relies on measuring each particle's mobility 
by observing configurations of particles undergoing rearrangements in either simulated or experimental systems. 
To accomplish this, the method requires a means to quantify the amount by which each particle participates in a given rearrangement.
This choice of dynamical characterization determines the type of statistical model that is most appropriate to
identify correlations between particle dynamics and local structure.
In this section, we describe the characterization of particle dynamics used by the original softness method 
and how it leads to a classification-based approach.
We then show how the original approach may be generalized to allow for a simpler regression-based model,
greatly improving its power when applied to limited data.
By choosing a more natural characterization, we are better able to take advantage of available data.

\subsection{Classification via Dynamical Outliers}

In order to define an observable proxy for particle mobility, we start with the critical mode $\vec{u}$ describing the onset of a rearrangement, as defined in the previous section.
Since rearrangements are characterized by the relative motion within local neighborhoods of particles, 
we use a measure of the local non-affine motion in each particle's environment, commonly referred to as $D^2_{\min}$~\cite{Falk1998}.
This ensures that particles which move together as rigid clusters are assigned small measures of motion, even if they display large displacements.
We define this quantity for a particle $i$ in the standard way,
\begin{align}
\dmin(i) = \min_{F}\qty{ \frac{1}{\abs{\mathcal{N}_i(\ell)}} \sum_{j\in \mathcal{N}_i(\ell)} \qty(F\Delta \vec{x}_{ij} - \Delta \vec{u}_{ij})^2},
\end{align}
where $F$ the deformation gradient matrix of size $d\times d$ calculated to minimize $\dmin(i)$.
The sum iterates over all particles in the neighborhood of particle $i$ (excluding itself), represented by the set $\mathcal{N}_i(\ell)$ with size $\abs{\mathcal{N}_i(\ell)}$.
The size of the neighborhood is set by a discrete cutoff distance $\ell$, 
which we calculate using the minimum path length between pairs of particles in the Delaunay triangulation of the configuration
(typically equivalent to a Euclidean distance of 2-3 particles diameters; see Sec.~\ref{sec:descriptors} for a comprehensive discussion).
The vector $\Delta \vec{x}_{ij} = \vec{x}_j - \vec{x}_i$ is the position of particle $j$ relative to $i$ 
and $\Delta \vec{u}_{ij} = \vec{u}_j - \vec{u}_i$ is the relative displacement between the two particles calculated from the critical mode.

\begin{figure}[t!]
\centering
\includegraphics[width=1.0\linewidth]{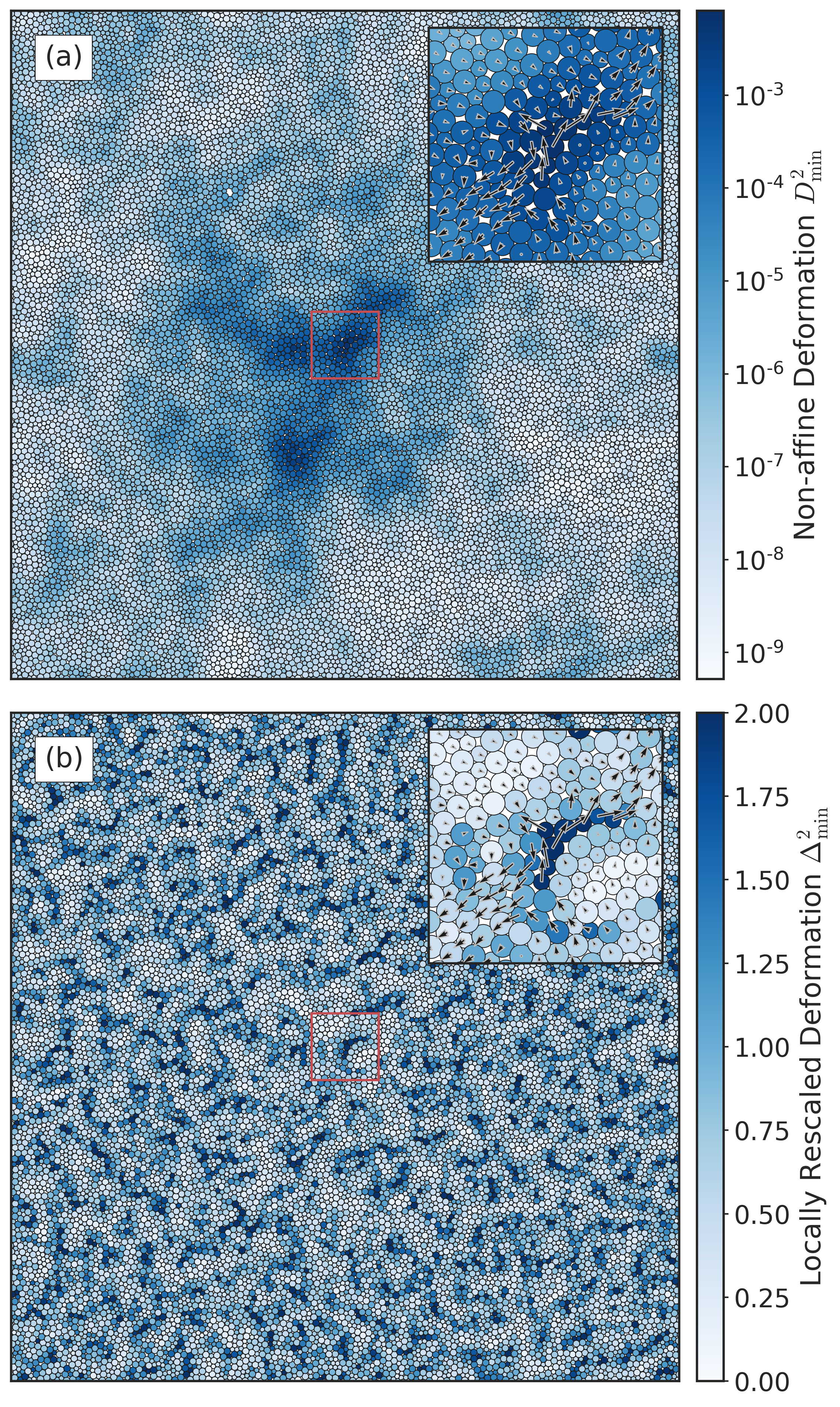}
\caption{
The onset of a rearrangement within a two-dimensional configuration of jammed soft particles undergoing shear.
(a) The non-affine deformation $\dmin$ for each particle in the configuration calculated from the critical mode $\vec{u}$, describing the onset of the rearrangement.
(Inset) Zoomed-in view of the neighborhood within the red box located near the particle with the largest value of $\dmin$, defining the source of the rearrangement. 
Arrows indicate each particle's motion within the critical mode.
(b) The locally rescaled non-affine deformation $\delmin$ calculated for each particle and (Inset) the corresponding zoomed-in view of the rearrangement source.
}
\label{fig:rearrangement}
\end{figure}

Fig.~\ref{fig:rearrangement}(a) depicts $\dmin$ for each particle in a configuration at the onset of a rearrangement.
The inset shows the local neighborhood at the ``source'' of the rearrangement 
(defined as the particle with largest value of $\dmin$) with arrows depicting the critical mode $\vec{u}$,
indicating the initial particle movements.
We expect $\vec{u}$ to decay like $r^{1-d}$ where $r$ is the distance from the source of the event and $d$ is the dimension~\cite{Picard2004}.
This suggests a power law dependence of $r^{-d}$ for for the non-affine motion as measured by $\dmin$, 
which depends on the difference in motion between adjacent particles and thus scales like the strain.

Once $\dmin$ has been calculated, the naive approach would be to perform a simple linear regression to determine the correlation between each particle's observed $\dmin$ and a measure of its local structure.
However, the power-law dependence of $\dmin$ poses a practical problem; 
it is both system-size dependent and ranges over many orders of magnitude for a given rearrangement 
(e.g. almost 10 orders of magnitude in the example depicted in Fig.~\ref{fig:rearrangement}(a)).
The result is that any correlations between $\dmin$ and any structural quantities of interest are weakened.
Indeed, we find that a linear model based on $\dmin$ and the various structural quantities we consider in this work only account for a small percentage of the observed variance.

To avoid this issue, the original softness method converts the problem into one of classification, 
defining two classes of particles: rearrangers, which will rearrange in the near future, and non-rearrangers, which have not rearranged for a long time (or strain in this case).
To train the classifier, examples of particles from both classes are identified by imposing strict cutoff thresholds on observed values of $\dmin$. 
In each configuration, examples of rearrangers are found by choosing particles such that $\dmin$ is greater than a cutoff $q_{\text{r}}$.
Similarly, examples of non-rearrangers are chosen by identifying particles with low observed $\dmin$ within a relatively long window of time into the future.
In this work, we set an upper threshold $q_{\text{nr}}$ on the maximum $\dmin$ experienced by a particle within a window of 10 rearrangement events into the future (including the current event).

This strategy reduces noise by limiting training to dynamical outliers, or particles in the tails of the distribution of $\dmin$, as it is much easier to distinguish between particles in these two regimes.
The primary problem with this approach is that there is not always a natural choice for the thresholds $q_{\text{r}}$ and $q_{\text{nr}}$.
As we will demonstrate, the choice of these cutoffs can dramatically affect the resulting classification accuracy of the model.
To the maximize the accuracy, these threshold must be taken to be very strict, greatly reducing the number of particles that can be utilized from each individual configuration.
As a result, achieving adequate accuracy with such a strategy would require a prohibitively large number of samples.
While loosening the strictness of the thresholds increases the number of samples,
it also introduces noise as some particles may have large values of $\dmin$ simply due to their proximity to the rearrangement, 
even though structurally they are indistinguishable from non-rearranging particles.
Similarly, particles with small values of $\dmin$ may simply be far away from the source of the rearrangement, but still have local structures with particularly low stability.

\subsection{Regression via Locally Rescaled Motion}

To remedy these problems, we take note of the fact that previous studies have determined that a typical jammed packing can have many soft spots, 
or local regions with particularly low energy barriers to rearrangement,
with the source of the rearrangement contained within just one (or a small number) of them~\citep{Patinet2016}.
Since a rearrangement in a homogeneous system would consist of a local plastic event surrounded by a decaying strain field described by the Eshelby kernel~\cite{Picard2004}, 
we hypothesize that the related critical mode will consist of such a field interacting with the underlying structural softness field.
The result of this interaction will manifest as small-scale variations on top of the continuum response.
In this view, soft spots are not associated simply with particles with relatively large values of $\dmin$, but rather with particles with large observed motion \textit{relative} to particles within their local environment. 
Therefore, we would like to normalize $\dmin$ so that it is independent of position relative to the the source of the rearrangement, 
but still captures particle level variations in the response.

To accomplish this, we introduce a simple modification to the $\dmin$ field.
For each particle within the $i$th particle's neighborhood including itself, we measure $\dmin(i)$ and calculate the average $\expval{\dmin}_{\mathcal{N}_i(\ell)}$.
We then rescale $\dmin(i)$ by this average value to obtain 
\begin{align}
\delmin(i) = \dmin(i) / \expval{\dmin}_{\mathcal{N}_i(\ell)}.
\end{align}
For simplicity we consider the same neighborhood of particles for both the original calculation of $\dmin$ and this locally rescaled version, but in principle, each could be chosen separately.
We choose the discrete cutoff distance $\ell$ such that it is large enough to capture local variations in the $\dmin$,
but not too large as to lose information about potential soft spots.
We find that a distance of $\ell=2$ is as small as possible while still capturing local variations in the critical mode,
commensurate with known length scales of $\dmin$ spatial correlations~\cite{Cubuk2017}.

Fig~\ref{fig:rearrangement}(b) depicts this locally rescaled measure of the non-affine deformation for the rearrangement in Fig~\ref{fig:rearrangement}(a). 
We see that $\delmin$ still captures local fluctuations in the response, 
but eliminates distance and angular dependencies without having to fit a functional form or explicitly address finite size effects.
We posit that this measurement of a particle's participation in a rearrangement
is a proxy for a particle's mobility.
Since $\delmin$ no longer varies over many orders of magnitude with distance from the rearrangement source,
standard linear regression now becomes practical. This change of training strategy allows us to avoid choosing cutoffs to identify training examples (rearrangers and non-rearrangers). Instead, we can utilize all of the particles in the system.
We will see that this dramatically improves predictive accuracy when data is limited.

\section{Structural Characterization of Local Particle Environment}

Now that we have defined a dynamical quantity that locally quantifies each particle's participation in a rearrangement, 
the next step is to characterize each particle's local structure.
Typically, a choice must be made as to which aspects of the local structure to measure. 
The original softness method uses structural descriptors derived from a particle's local pair correlation function~\citep{Cubuk2015}.
These structural descriptors were originally proposed by Behler and Parrinello as a means to parameterize potential energy surfaces for use in density-functional theory~\cite{Behler2007}.
In effect, the Behler-Parrinello  (BP) descriptors form an arbitrary basis which provides an over-determined representation of local structure (see Appendix~\ref{sec:BP}).
Because they are not specialized for any particular system,
the number of necessary descriptors can be very large, 
even after many redundant or non-informative features have been eliminated by the training process.
This means that the resulting form of softness, composed of a linear combination of these parameters, can be difficult to interpret.

Rather than choose an arbitrary basis of descriptors,
here we turn to persistent homology, a method in topological data analysis, to systematically identify a natural set of descriptors.
This procedure both minimizes guesswork and provides descriptors that are both tailored to the system of interest and easier to interpret.
In the past, the persistence algorithm has been used to study various topological aspects of configurations of particles in two-dimensions and higher~\cite{Kramar2013, Kramar2014, Hiraoka2016}.
In this section, we outline the procedure for applying the persistence algorithm to jammed packings of particles.
We then measure the statistical properties of the topological features within such systems
and explain their physical interpretations.

\subsection{Persistent Homology}

Persistent homology is a technique that detects and characterizes topological features contained within geometrically and/or topologically structured data~\cite{Edelsbrunner2010, Otter2017}.
In this case, we use it to characterize each two-dimensional configuration of jammed soft spheres at the onset of rearrangement.
For each particle $i$ we know its position $\vec{x}_i$ and its interaction radius $R_i$. 
In general, the types of topological features the persistence algorithm can detect include connected components, loops, voids, etc.
While the first two types of features are relevant in two dimensions, we primarily focus on loops in this study 
(or one-dimensional cycles, which we refer to simply as cycles from now on).

\begin{figure}[t!]
\centering
\includegraphics[width=0.95\linewidth]{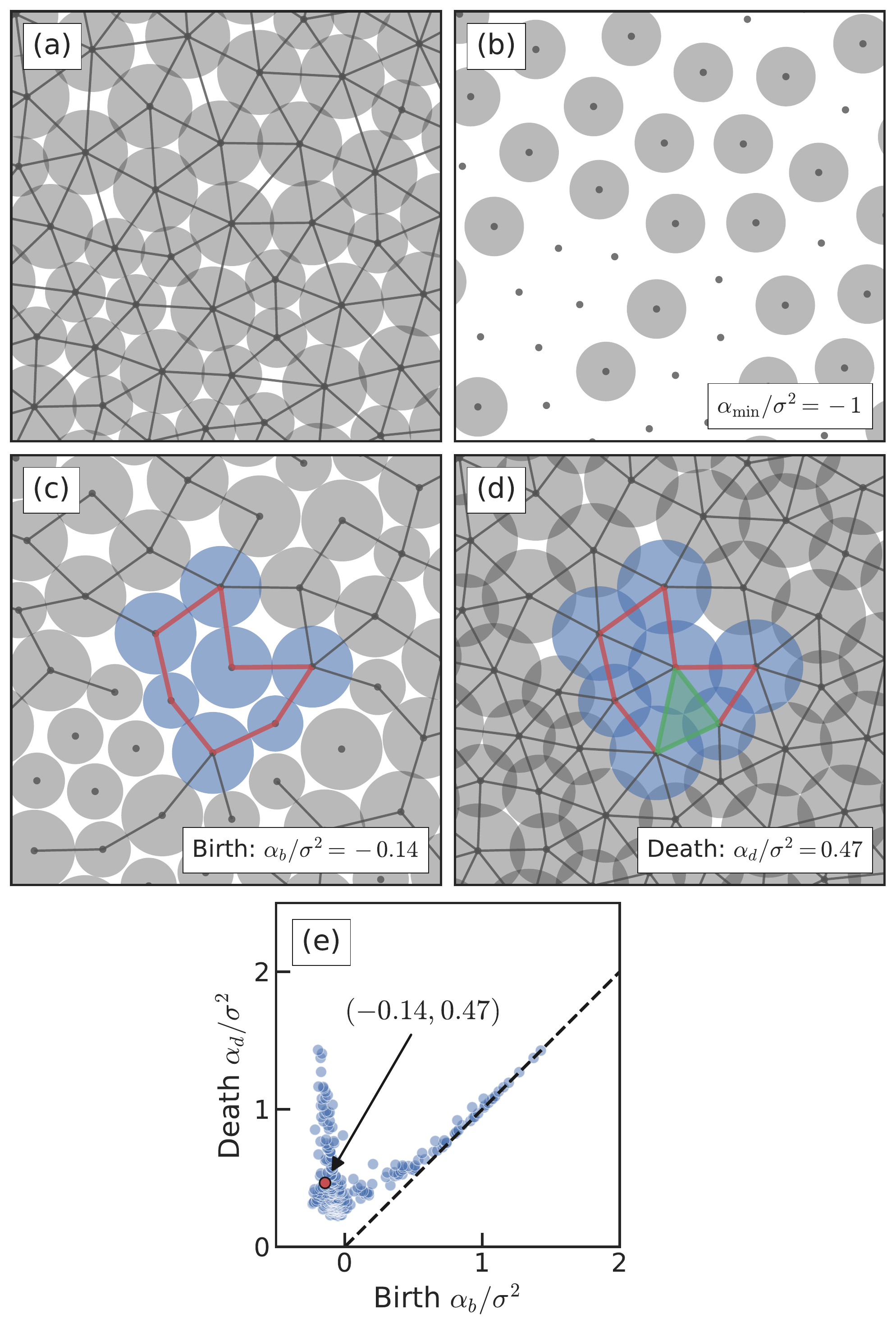}
\caption{(a) Weighted Delaunay triangulation of a packing of particles consisting of vertices at the center of each particle,
edges between neighboring particles, and triangles between triplets of mutually adjacent particles. 
Some edges in the triangulation correspond to particle contacts, while others do not.
(b) Initial configuration encountered during the filtration at $\alpha_{\min} = -\sigma^2$. 
Particles with radii $\sigma$ first appear as points,
while particles with larger radii begin as as finite-sized discs. 
The corresponding alpha complex, a subset of the Delaunay triangulation, consists of a single point at the center of every particle.
(c) Birth of the cycle consisting of the red edges at $\alpha_b = -0.014\sigma^2$, representing the overlaps between the discs highlighted in blue. 
The current alpha complex consists of the vertices and edges shown as black lines.
(d) Death of the cycle from (c) at  $\alpha_d = -0.014 \sigma^2$ when the green triangle is placed in the triangulation, representing the mutual overlap of the discs at its corners.
The alpha complex has additional edges compared to (c), along with triangles wherever three discs overlap (triangles not shown).
(e) Resulting persistence diagram quantifying all cycles encountered in the configuration. 
The cycle that is born in (c) and dies in (d) is highlighted in red.}
\label{fig:filtration}
\end{figure}

To perform the persistence algorithm on a configuration of particles, we perform a filtration of its weighted Delaunay triangulation~\cite{Edelsbrunner2010}.
To apply this filtration, we start by calculating the weighted Delaunay triangulation of the configuration of particles, using the squared radius $R_i^2$ of each particle as its weight.
Fig.~\ref{fig:filtration}(a) depicts a configuration and its associated weighted Delaunay triangulation.
This triangulation has the property that each contact between a pair of particles correspond to an edge in the triangulation (although the converse is not always true).
This ensures that it encodes the particle contact topology and therefore the mathematical constraints of the system.
In two dimensions, this Delaunay triangulation is composed of three different types of simplices: vertices, edges, and triangles (in three dimensions we would also have tetrahedra).
The filtration assigns an ordering to each of these elements from which we can build up the triangulation piece by piece.
The subsets of the triangulation we observe at each step are called alpha complexes, 
providing representations of the configuration at different length scales until we achieve the full Delaunay triangulation.

To find the ordering of simplices, we place a disc (or $d$-dimensional ball in $d$ dimensions)  of radius 
\begin{align}
r_i(\alpha) = \sqrt{ R_i^2 + \alpha}
\end{align}
at the center of each particle $i$.
Next, we use the control parameter $\alpha$ to gradually increase the size of these discs.
At the value $\alpha = 0$, each disc has the same radius as its corresponding particle in the configuration,
while for $\alpha < 0$ ($>0$) each disc is smaller (larger) than its corresponding particle.
Initially, we start with a value of $\alpha = -\sigma^2$ such that none of the discs overlap, where $\sigma$ is the minimum interaction radius of all the particles.
As shown in Fig.~\ref{fig:filtration}(b), particles with radii of $\sigma$ initially appear as points, while particles with larger radii are finite discs.
Each point or disc represents a separate connected component corresponding to a single vertex in the Delaunay triangulation.

At this point, the alpha complex consists of all the vertices, but none of the edges nor triangles.
As $\alpha$ increases, we consider the union of the discs; 
if a pair of discs starts to overlap and there exists an edge between the corresponding vertices in the full Delaunay triangulation, we add the edge to the alpha complex.
Similarly, at the instant that a triplet of discs starts to overlap at a single point 
and there exists a triangle composed of the associated vertices in the Delaunay triangulation, we add the triangle to the alpha complex.

If enough edges have been added, a cycle of edges may appear surrounding a hole in the union of discs.
When this occurs, we say that the cycle is ``born'' and record the value of $\alpha$ at that instance, $\alpha_b$.
Fig.~\ref{fig:filtration}(c) shows the birth of a new cycle at $\alpha_b = -0.014 \sigma^2$ highlighted in red with the participating discs in blue.
As $\alpha$ further increases, the hole that the cycle surrounds can break up into smaller holes as edges are added and shrink as triangles are placed in the alpha complex.
Eventually, when a hole is completely filled in we say the corresponding cycle has ``died'' and again record the value of $\alpha$ for this event, $\alpha_d$. 
Fig.~\ref{fig:filtration}(d) shows the death of the red cycle at $\alpha_d = 0.47 \sigma^2$.
At this instance, the triangle highlighted in green is placed into the alpha complex, plugging the hole that the cycle surrounds.

The value $\alpha_b$ for each cycle measures the length of the largest edge comprising that cycle, 
while $\alpha_d$ measures the overall scale of the cycle. 
We continue increasing $\alpha$ until the discs fill all of space and the Delaunay triangulation is complete.
In this way, each cycle that appears during the filtration is assigned a birth-death pair $(\alpha_b, \alpha_d)$ encoding its inherent length scales.
We plot this birth-death pair on a persistence diagram as demonstrated in Fig.~\ref{fig:filtration}(e). 
The collection of all birth-death pairs encodes the complete topological information at all length scales contained within the configuration.
For a more detailed mathematical explanation of the persistence algorithm, we refer the reader to Ref.~\onlinecite{Edelsbrunner2010}.
We generate weighted Delaunay triangulations using CGAL~\cite{cgal}. 
We also note that CGAL can be used to compute $\alpha$-values and the associated filtrations,
although we used our own implementation.

\subsection{Topological Structure of Jammed Packings}

\begin{figure*}[t!]
\centering
\includegraphics[width=1.0\linewidth]{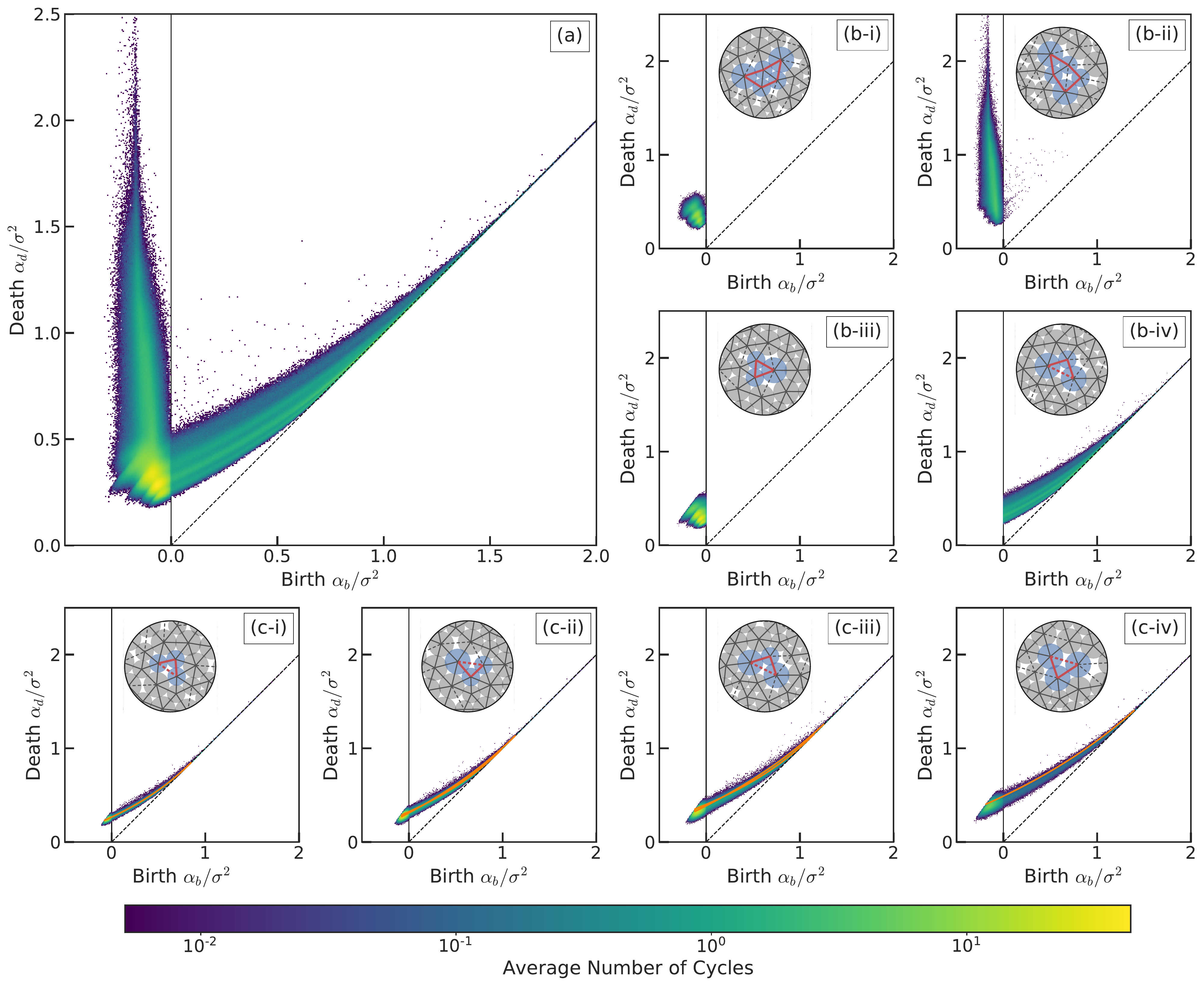}
\caption{(a) Average persistence diagram calculated from configurations at the onset of rearranging. 
Each pixel represents the the number of cycle observed at that particle $\alpha_b$ and $\alpha_d$ divided by the total number of configurations examined.
The black vertical line highlights $\alpha_b=0$, while the black dashed line highlights $\alpha_b=\alpha_d$.
(b) Decomposition of the persistence diagram into cycles of different sizes and edge types.
The vertical band is composed of (b-i) cycles with more than three edges and no interior gaps
and (b-ii) cycles with more than three edges and at least one interior gap,
while the diagonal band is composed (b-iii) cycles with exactly three edges and no gaps
and (b-iv) cycles with three edges and at least one gap.
(c) Decomposition of the diagonal band consisting of cycles with three edges according to constituent particle species.
The sub-bands are composed of triangular cycles with (c-i) three small particles, (c-ii) two small particles and one large particle,
(c-iii) one small particle and two large particles, and (c-iv) three large particles.
Analytical predictions of the four sub-bands are shown as orange curves (see Appendix, Sec.~\ref{sec:tricurves}).
Examples of each cycle type are shown in the insets, highlighted by the red edges and blue discs.
Contacts are depicted as solid lines, while gaps are dashed lines.
}
\label{fig:pdiag}
\end{figure*}

We use the persistence algorithm to analyze the topological structure of each of our rearrangement configurations.
For each configuration, we sort cycles into different bins according to their birth and death values and counting the number of cycles in each bin.
We then average the multiplicity across each configuration.
Fig.~\ref{fig:pdiag}(a) depicts this composite persistence diagram of one-dimensional cycles with each bin represented by a pixel.
To aid the eye, we have placed a solid black vertical line at $\alpha_b = 0$, along with a black dashed diagonal line along  $\alpha_b = \alpha_d$.
We observe two distinct bands of features in the persistence diagram: 
one located at negative $\alpha_b$, spanning a range of $\alpha_d$,
and a second which runs directly above the diagonal line where $\alpha_b = \alpha_d$, moving closer to this line as $\alpha_b$ increases.
These two bands meet in the lower left-hand corner at negative $\alpha_b$ and small $\alpha_d$, resulting in a highly concentrated set of peaks.
The diagonal band is composed of a set of sub-bands which each end on one of these peaks.

All the characteristics of the persistence diagrams we have noted correspond to different types of prominent features present in the particle configurations --  
in this case one-dimensional cycles.
In order to interpret the precise meanings of these features, we identify and classify each cycle according to both its size and composition.
The size is determined by the number of particles, or equivalently, the number of edges in the underlying Delaunay triangulation.
Composition is determined by the types (radii) of the particles involved, along with the types of the edges.
In the original configuration (corresponding to $\alpha=0$ in the filtration where each disc is the same radius as its particle), 
an edge corresponds to two particles which either overlap or do not overlap, which we call \textit{contacts} and \textit{gaps}, respectively.
Cycles can be composed of any combination of contacts and gaps.
If a cycle is born with $\alpha_b < 0$, then its largest length edge corresponds to a contact and the cycle must therefore be completely composed of contacts.
Conversely, if a cycle is born with $\alpha_b > 0$, its largest edge is a gap, but it may also contain some contacts.
Finally, if a cycle has more than three edges, we can assess whether it contains any gaps in its interior.
In two-dimensions, each cycle is the boundary of a two-dimensional surface composed of triangles. 
An interior edge is one contained in this surface, but not located on its boundary.

Since the persistence algorithm does not provide a unique representation of the cycles it detects, 
we choose a particular representation of each cycle when it is born.
We explain our procedure for identifying these birth cycles in the Appendix in Sec.~\ref{sec:repcycles}.
However, the exact choice we make does not affect the overall results.
Once we have associated each point in the persistence diagram with a particular cycle,
we see that specific types of cycles have births/deaths in different regions of the persistence diagram.
In Figs.~\ref{fig:pdiag}(b-i)-(b-iv), we sort cycles according to size and composition in terms of contacts and gaps.
Inset within each panel is a representative example of the type of cycle observed in that region of the persistence diagram.
As shown in Figs.~\ref{fig:pdiag}(b-i) and (b-ii), cycles with more than three edges are located all throughout the vertical band.
Cycles which contain at least one gap in their interior are located in the upper part, 
while those without interior gaps concentrate in the lower part.
Fig.~\ref{fig:pdiag}(b-iii) shows that the lower part of the band also contains triangular cycles, containing exactly three contacts.
Since the vertical band is located at $\alpha_b < 0$, all of these cycles are composed solely of contacts.
Depicted in Fig.~\ref{fig:pdiag}(B-iv), the band running along the diagonal also contains triangular cycles composed of exactly three edges, 
coinciding with and extending out from the lower part of the vertical band.
When $\alpha_b >0$, these triangular cycles will always contain one or more gaps.

\begin{figure*}[t!]
\centering
\includegraphics[width=1.0\linewidth]{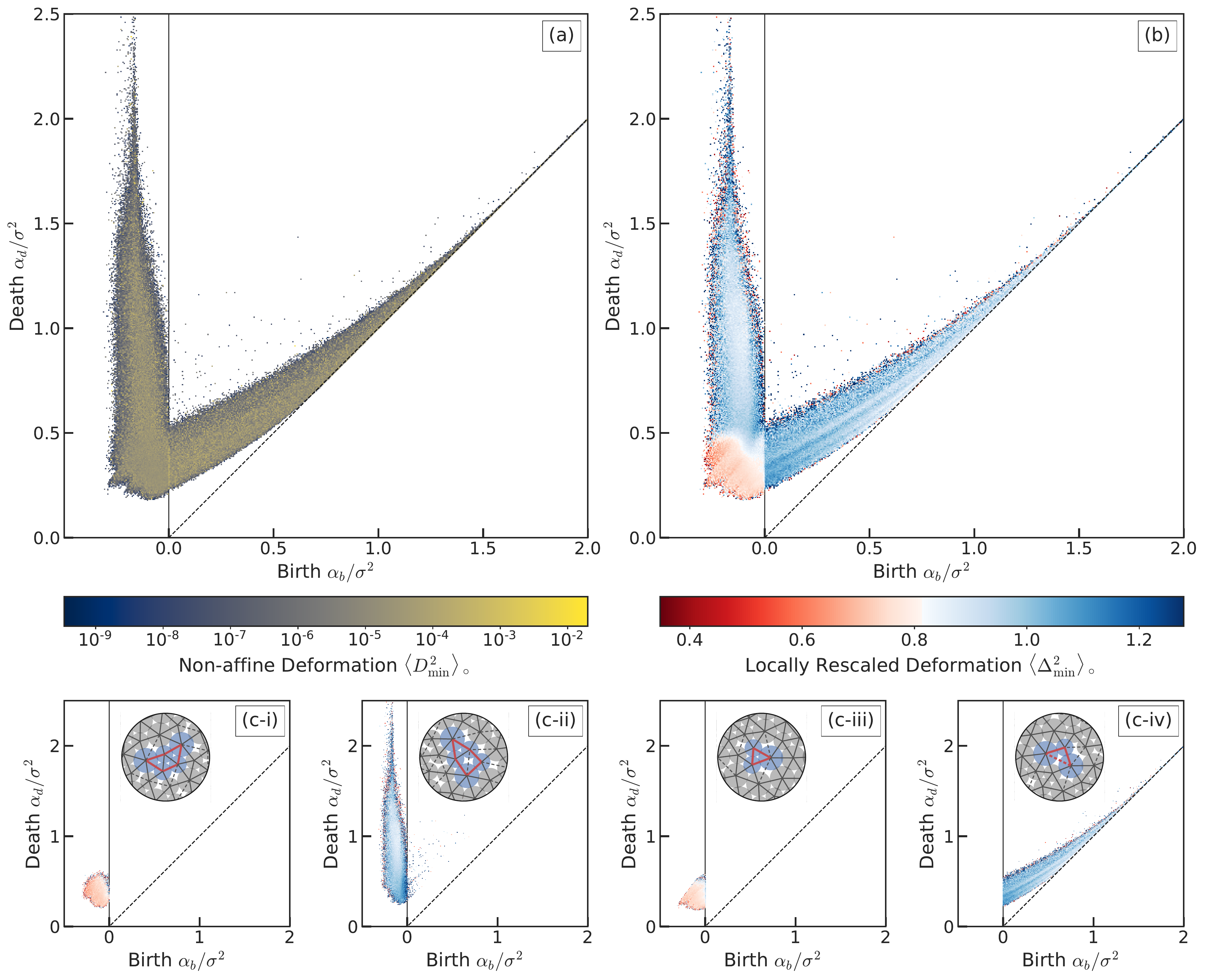}
\caption{(a) Correlation between cycles detected by the persistence algorithm and the non-affine motion of the particles comprising each cycle.
Each pixel is colored according to the value of $\langle \dmin \rangle_\circ$ averaged across all cycles in that bin.
There is no significant correlation between the persistence diagram and $\langle \dmin \rangle_\circ$
besides a small excess of motion for cycles exactly at $\alpha_b = 0$.
The black vertical line highlights $\alpha_b=0$, while the black dashed line highlights $\alpha_b=\alpha_d$.
(b) Correlation between cycles and their normalized motion $\langle \delmin \rangle_\circ$ averaged across all cycles within each bin.
There are significant correlations between regions of the persistence diagram and $\langle \delmin \rangle_\circ$.
(c) Decomposition of the persistence diagram according to number of edges and edge types correlated with $\delmin$.
Cycles with small values of $\delmin$ (shown in red) tend to consist of (c-i) more than three edges and contain no interior gaps
or (c-iii) exactly three edges with no gaps.
Cycles with large values of $\delmin$ (shown in blue) tend to consist of (c-ii) more than three edges and at least one interior gap
or (c-iv) exactly three edges with at least one gap.
Examples of each cycle type are shown in the insets, highlighted by the red edges and blue discs.
Contacts are depicted as solid lines, while gaps are dashed lines.
}

\label{fig:pdiag_D2min_normed}
\end{figure*}

The decomposition of the persistence diagram can be taken one step further to understand the effects of particle size and position on the triangular cycles.
In Figs.~\ref{fig:pdiag}(C-i)-(C-iv), we have sorted the triangular cycles into groups based on the combinations of particle sizes.
Again, inset within each panel is a representative example of a triangular cycle in that region of the persistence diagram.
Since there are two possible particle radii, we observe four different combinations of three particles: 
(c-i) three small particles, (c-ii) two small particles and one large particle, (c-iii) one small particle and two large particles, and (c-iv) three large particles.

For each of these cases, we can analytically calculate a birth-death curve which approximates the sub-band, highlighted by the orange curves.
Starting with three particles arranged into a triangle with three contacts, 
we calculate $\alpha_b$ and $\alpha_d$ as we continuously open up one of the contacts into a gap.
We explain this calculation in more detail in the Appendix in  Sec.~\ref{sec:tricurves}.
As the gap opens up and a triangle becomes more elongated and less regular, both $\alpha_b$ and $\alpha_d$ increase while the difference between them decreases.
Eventually, the triangle elongates so much that $\alpha_b = \alpha_d$ and the curve ends at its intersection with the diagonal line.
Each of the different combinations of particles comprises a separate curve. 
In the cases with one large particle and two small particles or one small particle and two large particles, 
there are two places where the gap can be placed: between particles of the same type or particles of differing types. 
This means we can calculate two different curves for each of these cases. 
However, these curves are so close together that it is difficult to distinguish between them within the corresponding sub-bands.
In addition, contacts in these cycles can have varying amounts of overlap and sometimes cycles can have more than one gap. 
Both of these effects contribute to the widths of the sub-bands.

In summary, calculating composite persistence diagrams for our jammed configurations provides a rigorous statistical representation of local topological structures.
By identifying the cycles corresponding to each point and then decomposing the persistence diagrams accordingly, we can fully understand how different types of cycles correspond to features we observe in the composite persistence diagram for all cycles.

\section{Connecting Dynamics and Structure}

Now that we have established quantitative descriptions of both a particle's dynamics during a rearrangement and its local structure,
we search for correlations between the two.
To do this, we color each pixel in the composite persistence diagrams according to the average amount of motion undergone by the cycles present in that pixel.
For each cycle, we measure value of either $\dmin$ or $\delmin$ averaged across all particles in that cycle, we we denote $\langle \dmin \rangle_\circ$ and $\langle \delmin \rangle_\circ$, respectively.
Next, we average each cycle-defined measure of dynamics for each pixel in the persistence diagrams across all cycles present in that pixel.

Fig.~\ref{fig:pdiag_D2min_normed}(a) shows the persistence diagram from Fig.~\ref{fig:pdiag}(a) correlated with $\dmin$, the non-normalized measure of non-affine deformation.
We see that $\langle \dmin \rangle_\circ$ is almost perfectly uniform across all regions of the persistence diagram, indicating no correlation between cycle type and $\dmin$. 
The only exception we observe is a very narrow vertical band at $\alpha_b=0$ which contains slightly larger measures of motion.
This indicates that particles that participate more in rearrangements tend to contain contacts that have such small numerical values of overlap that they are almost gaps.
However, this signal is very weak.

In contrast, Fig.~\ref{fig:pdiag_D2min_normed}(b) depicts the persistence diagram correlated with $\delmin$, the normalized measure of motion.
Here we observe very strong correlations between motion and cycle type; 
cycles located in the lower left-hand corner are typically located in neighborhoods with relatively low amounts of motion relative to their surroundings, 
while cycles located in either of the two bands tend to participate more strongly in rearrangements.
This contrast is especially strong in the diagonal band with an almost step-like jump in $\langle \delmin \rangle_\circ$ occurring across the $\alpha_b=0$ line.

In Figs.~\ref{fig:pdiag_D2min_normed}(c-i)-(c-iv), we correlate $\delmin$ with the persistence diagrams of the four classes of cycles. 
Insets depict examples of the types of cycles represented in each panel.
We see in Figs.~\ref{fig:pdiag_D2min_normed}(c-i) and (c-iii) that cycles with more than three edges and no interior gaps, 
along with triangles with no gaps, typically have low $\delmin$, colored in red.
On the other hand, Figs.~\ref{fig:pdiag_D2min_normed}(c-ii) and (c-iv) show that cycles with more than three edges that contain interior gaps, 
along with triangles that contain at least one gap,  typically have high $\delmin$, colored in blue.
This correspondence between gaps and $\delmin$ is also present in the sub-bands comprising the full diagonal band. 
The sub-band curves we show in Figs.~\ref{fig:pdiag}(c-i)-(c-iv) correspond to triangles with exactly one gap.
If a triangle has more than one gap, it will result in a larger $\alpha_d$, 
moving the cycle upwards in the persistence diagram away from the associated curve.
We see in Fig.~\ref{fig:pdiag_D2min_normed}(b) that the regions of persistence diagrams corresponding to triangles 
with more than one gap are a darker blue than those with one gap, indicating larger values of $\delmin$.

From all of these observations, we posit that a particle's participation in a rearrangement relative to its local environment, 
as measured by $\delmin$, is determined by the presence or absence of gaps, or conversely, the number of contacts. 
The more gaps, or less contacts, a particle shares with its nearest neighbors, 
the larger its participation in a given rearrangement will be relative to its local neighborhood.

\subsection{Topologically-Informed Structural Descriptors}\label{sec:descriptors}

\begin{figure}[b!]
\centering
\includegraphics[width=1.0\linewidth]{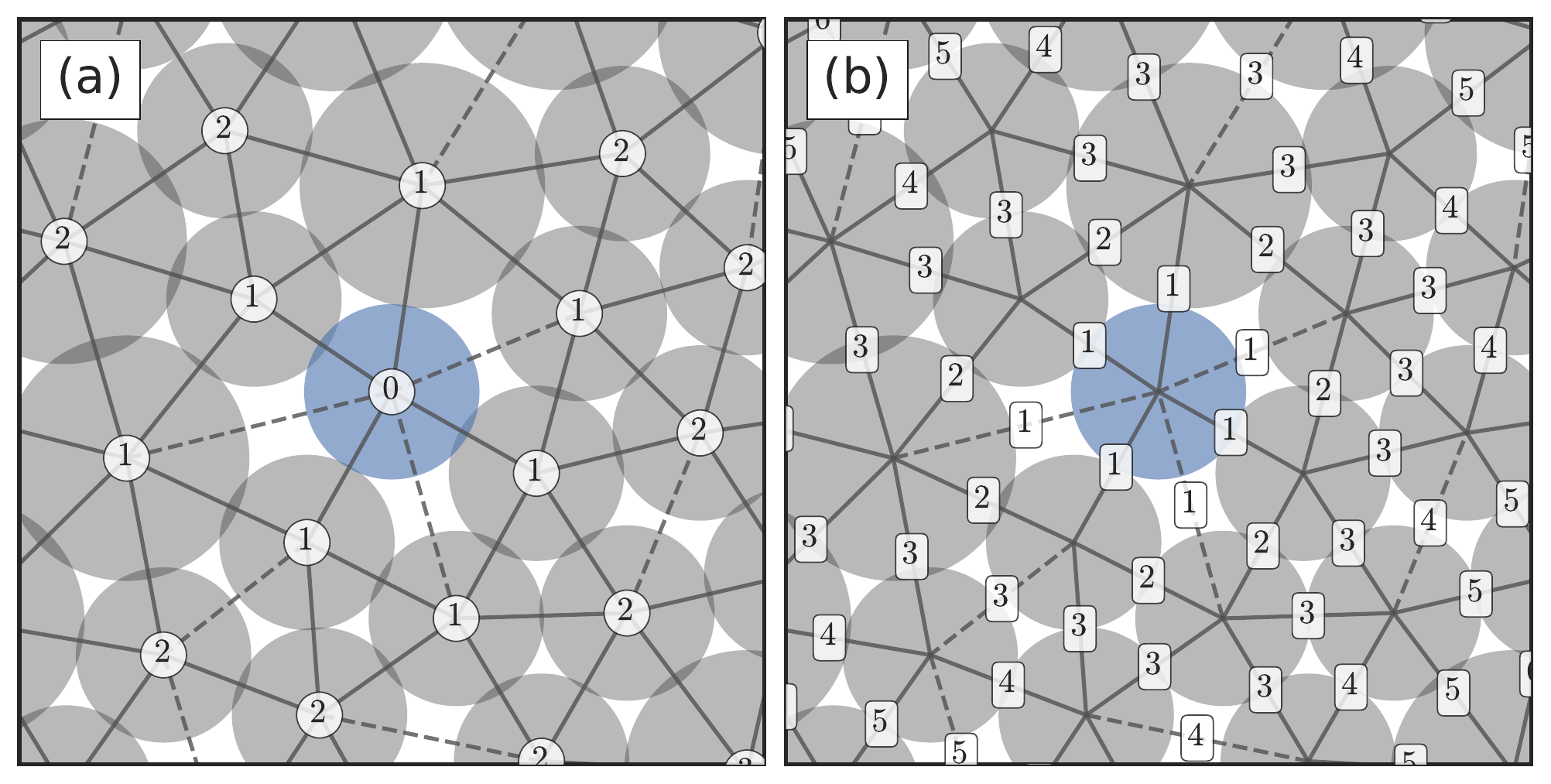}
\caption{Discrete distances defined in terms of a configuration's weighted Delaunay triangulation.
(a) Distance $d_{i, j}$ of each particle $j$ from the central particle $i$ shown in blue. 
Particle distances are taken as the minimum path length along the edges of the the triangulation between the two particles.
(b) Distance $d_{i, (j, k)}$ of each edge $(j, k)$ composed of particles $j$ and $k$ from the central particle $i$. 
Edge distances are calculated from the sum of the distances of their respective vertices from the particle of interest.}
\label{fig:tri_dist}
\end{figure}

Based on these observations, we use the numbers of gaps and contacts in a particle's local environment to construct a set of local structural descriptors.
In order to allow for the possibility that a particle is affected by more than just its immediate nearest neighbor structure,
we allow structural descriptors to be defined at a range of distances from a particle of interest.
Since gaps and contacts are defined in terms of the Delaunay triangulation which captures a configuration's contact structure,
we define all distances in terms of this triangulation.
We start by defining the distance $d_{jk}$ as the minimum path length in the triangulation between particles $j$ and $k$, 
counted in terms of the number of edges (e.g., nearest neighbors are distance one, next nearest neighbors are distance two, etc.).
Fig.~\ref{fig:tri_dist}(a) shows an example of these distances for a neighborhood around a specific particle shown in blue.
This measure of distance has the nice property that it is defined in a way which takes into account the contact topology, along with differences in particle radii.
We use this discrete distance in all aspects of our softness procedure, 
including the cutoff distance $\ell$ used to calculate $\dmin$ and $\delmin$ described previously.
In those cases, we consider all particles to be in the neighborhood of a particle $i$ if $d_{ij} \leq \ell$.

Next, we assign a measure of distance between a particle $i$ and particular edge $(j, k)$ defined in terms of its pair of vertices $j$ and $k$ as
\begin{align}
d_{i, (j, k)} =  d_{ij} + d_{ik},\label{eq:dist}
\end{align}
the sum of the distances of particles $j$ and $k$ from $i$.
As depicted in Fig.~\ref{fig:tri_dist}(b), this definition of distance separates the edges in the triangulation into ``layers'' at different distances.
Edges that are incident with particle $i$ are assigned distance $d_{i, (j, k)} = 1$,
while those that are incident with two nearest neighbors of $i$ are at distance $d_{i, (j, k)} = 2$, etc.

Using this definition of distance, we can simply count the number of gaps and contacts present in each layer of the Delaunay triangulation.
For a particle $i$, we denote the number of gaps and contacts located at a distance $d_{i, (j, k)} = m$ as $g_i^m$ and $c_i^m$, respectively.
We also include the particle species $p_i$, where $p_i=0$ if the radius $R_i = \sigma$ and $p_i=1$ otherwise.
The result is a list of structural descriptors 
\begin{align}
x^i = (p_i, g_i^1, c_i^1, \ldots, g_i^{\ell_{\max}}, c_i^{\ell_{\max}}) \label{eq:descriptors}
\end{align}
where $\ell_{\max}$ is the maximum distance considered.

\begin{figure}[t!]
\centering
\includegraphics[width=0.95\linewidth]{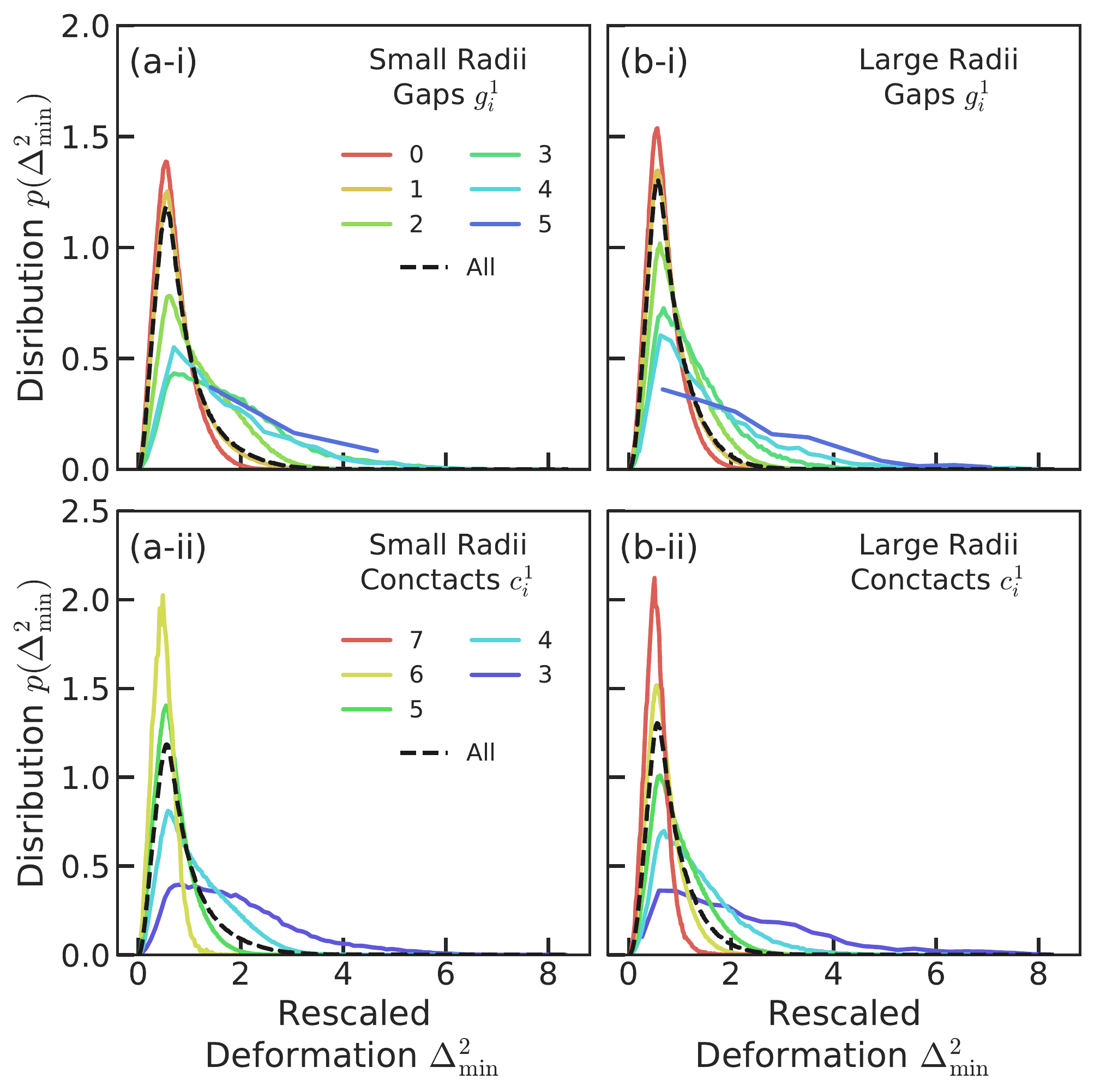}
\caption{Distributions of the locally rescaled non-affine deformation $\delmin$ for (a-i) small ($R_i=\sigma$) and (a-ii) large ($R_i=1.4\sigma$ particles with different numbers of gaps in the first layer of their local Delaunay triangulation $g_i^1$. 
Similarly, the distributions of $\delmin$ for (b-i) small and (b-ii) large ) particles with different numbers of contacts in their first layer $c_i^1$.
The full distributions including all particles are shown as black dashed curves.
Particles with more gaps or less contacts tend to have larger values of $\delmin$ on average.}
\label{fig:gc_dist}
\end{figure}

In Fig.~\ref{fig:gc_dist},  we plot the distributions of $\delmin$ 
for particles with different numbers of (a) gaps and (b) contacts in their nearest neighbor environments, $g_i^1$ and $c_i^1$, respectively.
Already, we see that the larger the number of gaps a particle has, and the lower its number of contacts, 
the larger value of $\delmin$ it will have on average.
To fully characterize this correlation, we perform linear regression with the structural descriptors $x_i$ acting as our independent variables 
and the locally rescaled non-affine deformation $\delmin$ acting as our dependent variable.
The result is a new definition of softness, composed of a linear combination of gaps and contacts at different discrete distances,
\begin{equation}
    S_i=\sum_\mu w_\mu x^i_\mu, \label{eq:Sdef}
\end{equation}
where $\mu$ is an index for the components of $x^i$ in Eq.~\ref{eq:descriptors} and the weights $w_\mu$ are determined by the regression.
In the next section, we compare this new formulation of softness with the previous version of the method.

\section{Results}\label{sec:results}

\begin{table*}[t!]
\setlength{\tabcolsep}{11pt}
\centering
\caption{\label{tab:compare}
Comparison of four different combinations of statistical model, dynamical measure and structural descriptors.}
\begin{ruledtabular}
\begin{tabular}{ccccc}
\thead{Model Type} & \thead{Dynamical\\Measure} & \thead{Structural\\Descriptors} & \thead{Accuracy\footnote{The accuracy metric is determined by model type: binary classification accuracy is used to measure classification success and $R^2$ is used to measure regression accuracy.
To compare all softness schemes, we report the percentile of the softness value of the particle with largest $\dmin$ averaged over each configuration, $\CDF$.} 
[\%]} & \thead{Global Maxima  Percentile\\$\CDF$ [\%]} \\ \hline
Classification & $\dmin$    & Behler-Parrinello   & $91.5\pm 1.1$     & $86.8\pm 0.8$    \\
Classification & $\dmin$    & Gaps/Contacts      & $96.7\pm 0.7$     & $89.1\pm 0.7$\\
Regression & $\delmin$ & Behler-Parrinello     & $18.51\pm 0.02$  & $88.7\pm 0.5$ \\
Regression & $\delmin$ & Gaps/Contacts          & $21.68\pm 0.03$  & $86.6\pm 0.7$\\
\end{tabular}
\end{ruledtabular}
\end{table*}

We separately compare each aspect of our new method with the previous version of softness. 
We test four different combinations of dynamical characterization ($\dmin$ or $\delmin$) and structural descriptors (BP descriptors or gaps/contacts).
The accuracy of each combination of methods is reported in Table~\ref{tab:compare}.
Results for jammed packings in higher spatial dimensions and for a variety of pressures are reported in Ref.~\onlinecite{Ridout2020}

When performing classification, we have chosen our cutoffs to identify examples of non-rearrangers and rearrangers, $q_{\text{nr}}$ and $q_{\text{r}}$,
as strictly as possible, limiting ourselves to one particle per class in each configuration.
These two particles exhibit the largest and smallest $\dmin$ in each configuration.
As we will demonstrate, this results in the highest possible classification accuracies when we use the SVM approach.

In all cases, we report a metric of accuracy appropriate to the type of model used.
For the classifation models, we report the binary classification accuracy, 
the percentage of particles that are correctly classified as having positive or negative softness.
For the regression-based models, we report $R^2$, the fraction of the variance in the dynamics explained by the model.

To ensure that we do not overfit our models, we perform cross-validation, 
training on one set of trajectories and computing test scores on another independent set of trajectories. 
When cross-validation demonstrates no significant difference between the training and testing accuracy,
 we report the mean and variance of the accuracy obtained via bootstrapping.
Otherwise, we report the mean and variance of the cross-validated test accuracy.
We have also chosen our model hyperparameters via cross-validation into order to maximize the accuracy.
We refer the reader to the Appendix, Sec.~\ref{sec:protocol}, for a complete description of our training procedures and choices of hyperparamters.

For the classification-based models using $\dmin$, we see that the structural descriptors based on gaps and contacts performs slightly better than the BP descriptors.
However, both sets of descriptors perform well with accuracies greater than $90\%$.
We see similar results for the regression-based schemes using $\delmin$, with gaps and contacts performing slightly better, but both sets of descriptors resulting in $R^2$ values around $20\%$.

We also see that a classification scheme based on $\delmin$ performs well for both sets of descriptors.
In contrast, performing regression on $\dmin$ performs very poorly, resulting in $R^2$ accuracies of only a couple percent.
This shows that $\delmin$ is a more robust measure of dynamics, performing well independent of the choice of statistical model.

We note that the classification accuracies in Table~\ref{tab:compare} tend to be much higher than the corresponding regression accuracies.
This is because classifiers only attempt to sort particles into binary classes 
and also only consider particles that could be considered as outliers in the distribution of $\dmin$ or $\delmin$.
This means that it is not appropriate to directly compare classification and regression accuracies. 
Since a sample may contain many ``soft spots,'' only one (or a few) of which will rearrange in a particular event,
a good criterion to measure success is whether or not the rearrangement always localizes around a soft particle. 
In order to compare both classes of models with this criterion, we follow the approach of Ref.~\onlinecite{Patinet2016}.
We identify the particle $i_{\max}$ in each configuration with largest $\dmin$, the global maximum.
This particle can be considered, in effect, the ``source'' of the rearrangement.
Next, we record the percentile of the global maximum's computed value of softness $S_{i_{\max}}$ within the distribution of all particles in its respective configuration.
This is equivalent to evaluating the cumulative distribution function of softness at $S_{i_{\max}}$, which we denote $\text{CDF}(S_{i_{\max}})$.
If a model were to perfectly predict which particle is most likely to rearrange within a configuration,
then the global maximum in $\dmin$ would coincide with the maximum value of softness in that configuration and we would obtain $\text{CDF}(S_{i_{\max}})=1$. If the model failed completely so that a random particle is chosen, then we would obtain $\text{CDF}(S_{i_{\max}})=0.5$.
We report $\CDF$, the average percentile of the global maxima in $\dmin$ in each configuration for all models in Table~\ref{tab:compare}.
 We find that all combinations of methods perform comparably well, 
consistently placing the global maxima in $\dmin$ in at least the $86$th percentile of softness.

\begin{figure}[b!]
\centering
\includegraphics[width=0.95\linewidth]{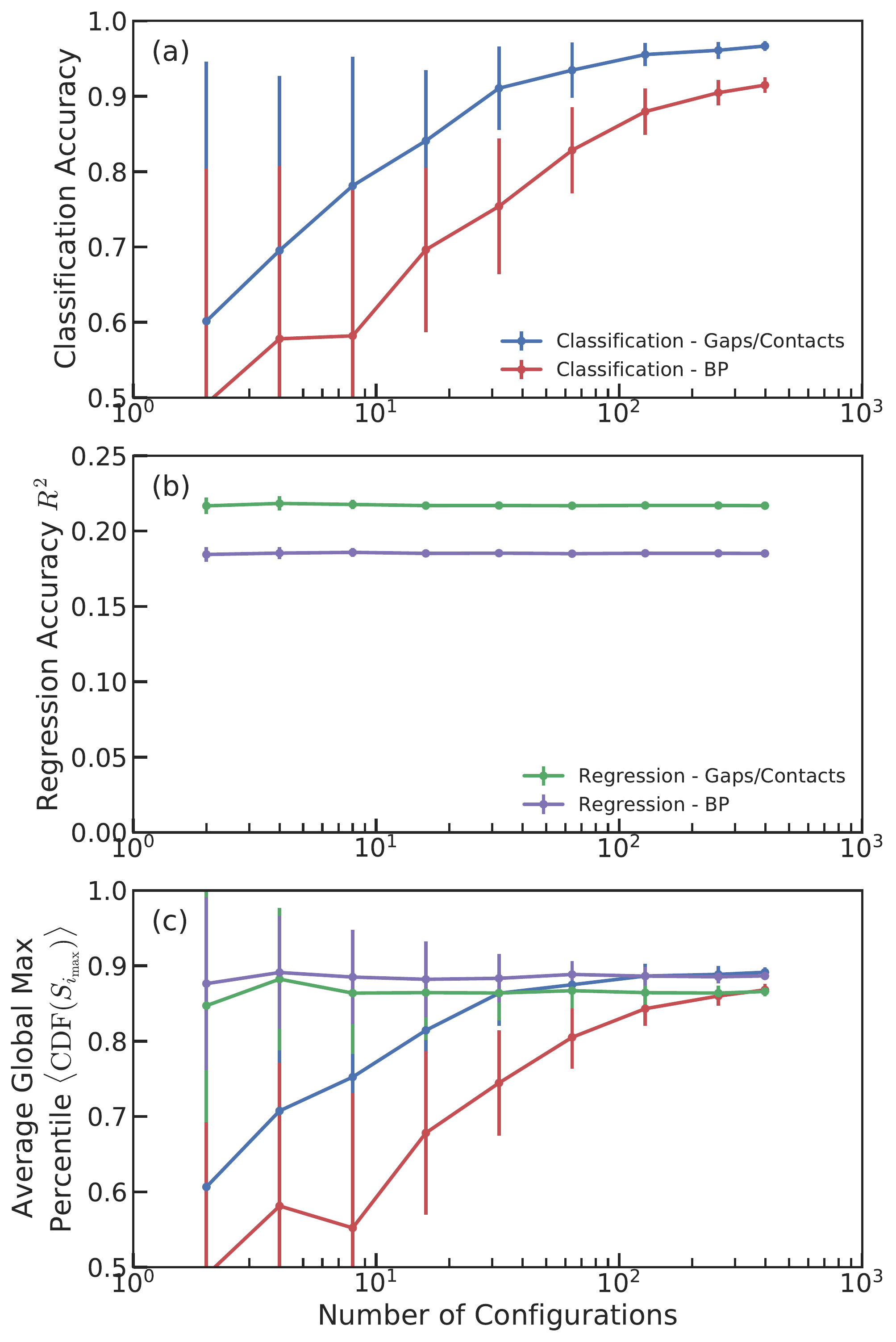}
\caption{
Model accuracies as a function of the number of independent configurations (one rearrangement configuration per trajectory) included in the training set.
(a) Classification accuracy using $\dmin$ with gaps and contacts (blue) and BP descriptors (red).
(b) Regression accuracy $R^2$ using $\delmin$ with gaps and contacts (green) and BP descriptors (magenta).
(c) Average percentile of the particle with largest $\dmin$ within each frame  for the four models, colored according to (a) and (b).
Error bars correspond to the variance of the accuracies computed via cross-validation or bootstrapping.
}
\label{fig:acc_ntraj}
\end{figure}

\begin{figure}[t!]
\centering
\includegraphics[width=0.95\linewidth]{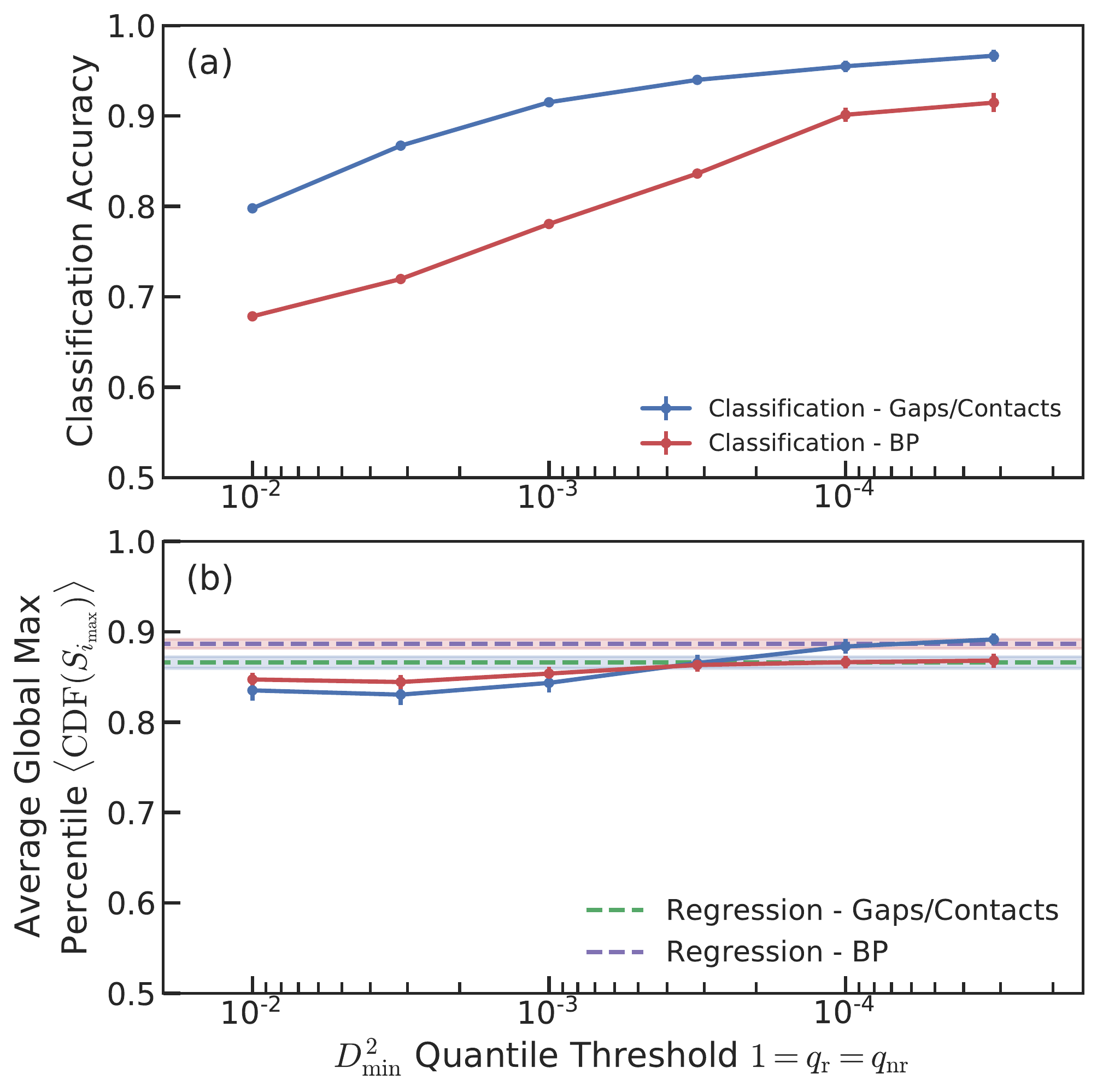}
\caption{Accuracies as a function of quantile thresholds $q=q_{\text{r}} = q_{\text{nr}}$ used to identify training examples of rearranging or non-rearranging particles for classification.
A smaller value of $q$ indicates a stricter threshold and smaller training set.
(a) Classification accuracy for a classifier trained using gaps and contacts (blue)  and  BP descriptors (red).
(b) Average percentile of the particle with largest $\dmin$ within each frame $\CDF$. 
The result for the classification models are are shown using solid lines while the corresponding accuracies for the regression models  using $\delmin$ are shown as dashed lines for gaps and contacts (green) and BP descriptors (magenta) .
Error bars correspond to the  variance of the accuracies computed via cross-validation or bootstrapping.
Similarly, transparent bands surrounding the dashed lines represent variance for the regression models.
}
\label{fig:acc_thresh}
\end{figure}

\begin{figure}[t!]
\centering
\includegraphics[width=0.95\linewidth]{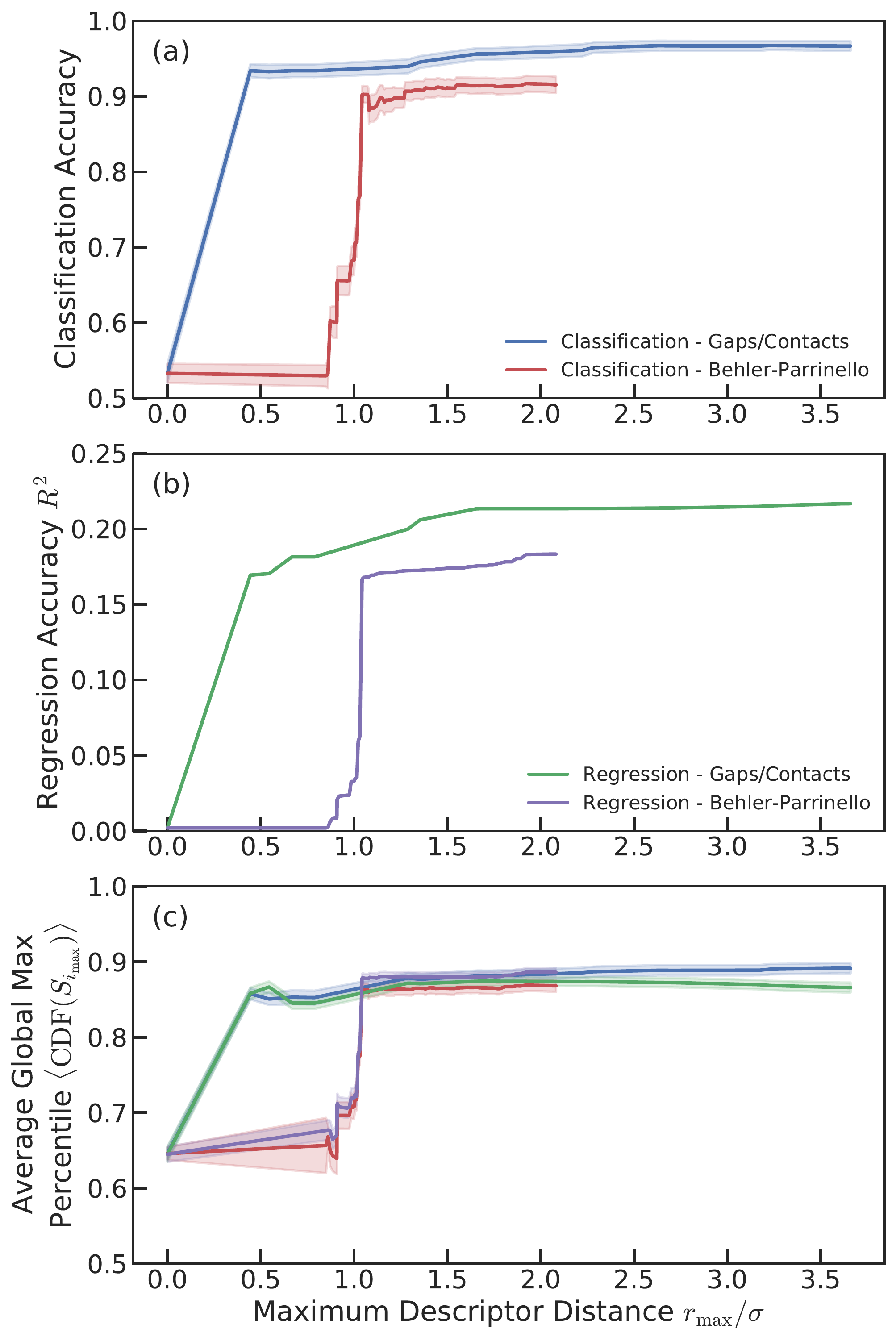}
\caption{Accuracies as a function of maximum Euclidean descriptor distance $r_{\max}$ for included structural descriptors. 
Distance for each descriptor is averaged over all instances of that feature and measured in units of $\sigma$, the minimum particle radius.
(a) Classification accuracies for the classification models using  gaps and contacts (blue) and  BP descriptors (red).
(b) Regression accuracies $R^2$ for regression models using gaps and contacts (green)and  BP descriptors (red).
(c)  Average percentile of the particle with largest $\dmin$ within each frame $\CDF$ for all four models.
Transparent bands surrounding the lines indicate  the variance of training accuracies computed via cross validation. }
\label{fig:acc_dist}
\end{figure}

One major benefit to using $\delmin$ with a regression-based scheme is the small amount of data needed to obtain high accuracy.
Fig.~\ref{fig:acc_ntraj} shows the accuracyof all four methods as a function of the number of configurations used in training. 
In order to calculate error bars via bootstrapping or cross-validation, the minimum number of trajectories needed is two.
In Fig.~\ref{fig:acc_ntraj}(a), we see that the classification accuracy is dramatically affected by the amount of training data for both sets of descriptor, and does not begin to level off until one has several hundred configurations.
In contrast, we see in Fig.~\ref{fig:acc_ntraj}(b) that the regression accuracy is already at its maximum when using the minimum number of configurations, namely 2.
In fact, a single rearrangement configuration would likely be sufficient to attain the maximum regression accuracy.
In Fig.~\ref{fig:acc_ntraj}(c), we plot $\CDF$ for the four different schemes.
Again, we see that in the case of classification, $\CDF$ is strongly dependent on the amount of training data,
while in the case of regression it is high even for two configurations. 
We note that while the variance of $\CDF$ is large for regression with small numbers of trajectories,
it decreases very quickly with additional data and the mean score remains high.

Another benefit to using regression in concert with $\delmin$ is that training examples of specific particles are not needed.
In Fig.~\ref{fig:acc_thresh}, we investigate the effect of the thresholds used to choose training examples for classification in combination with $\dmin$.
We parameterize these thresholds, $q_{\text{r}}$ and $q_{\text{nr}}$ in terms of the percentiles in the $\dmin$ distribution within each configuration.
For simplicity we choose to set these thresholds equal such that $q=q_{\text{r}} = q_{\text{nr}}$.
In Fig.~\ref{fig:acc_thresh}(a), we report the classification accuracy as a function of $q$.
We see that the accuracy is greatly affected by the choice of $q$.
In Fig.~\ref{fig:acc_thresh}(b) we observe that while $\CDF$ is far less sensitive,
it is still affected by the choice of $q$.
For reference, we have also provided the corresponding regression accuracies using $\delmin$ for both sets of descriptors, shown as dashed horizontal lines.
We see that $\CDF$ converges to these values at very strict thresholds.

Finally, we compare the dependence of the four schemes on the number of included structural descriptors and the size of the local environment that they encompass.
For both sets of descriptors, we sort each descriptor by its average Euclidean distance from the particle of interest (see Appendix, Sec.~\ref{sec:descriptors} for details).
Starting with the complete set of descriptors, we iteratively remove the descriptor at the largest distance, retraining our models at each step and measuring the new accuracies.
Fig.~\ref{fig:acc_dist} shows the training accuracies for our different models as a function of maximum descriptor distance $r_{\max}$.
In Figs.~\ref{fig:acc_dist}(a) and (b) we plot the accuracies for our classification and regression models, respectively, using both sets of descriptors. 
In Fig.~\ref{fig:acc_dist}(c) we compare $\CDF$ for all four methods.
We see that in all four cases the accuracy rapidly increases up to a distance of about $0.5$ to $1.0$ particle radii for both sets of descriptors,
with marginal improvements at larger separations.
For gaps and contacts, we find that only two descriptors are sufficient for a particle $i$:  the particle species $p_i$ the number of contacts of the particle with its neighbors $c_i^1$.
In contrast, the BP descriptors require at least around 20 descriptors to describe this environment. 
In principle, a different choice of parameters in the definitions of the descriptors could reduce this number,
but it is not clear what these parameters should be {\it a priori}.
We conclude that gaps and contacts provide a more concise description of the local structure.

\section{Discussion}

In summary, we have introduced two major improvements to the softness method.
First, we have defined a locally rescaled version of $\dmin$ -- represented as $\delmin$ -- which captures local variations in the relative motion of particles during rearrangements,
converting the softness problem from one of classification to regression.
The result is a more natural characterization which avoids the need to define classes of particles that are more or less likely to rearrange.
This allows us to take advantage of all particles in a given data set, rather than just statistical outliers in observed mobility, 
greatly reducing the number of configurations needed to compute softness. In our case, this improvement leads to a several-hundred-fold decrease in the amount of data needed to achieve an accuracy of $\CDF=0.86$. This is clearly a major advantage when there is limited data available for training, as is often the case in laboratory experiments.

Second, we have demonstrated a procedure for characterizing the local structure of particle configurations.
This procedure, based on persistent homology, allows for the systematic development of topologically-informed structural descriptors
that can be specialized to a system of interest.
This results in a more concise and interpretable set of descriptors 
which further decreases the amount of data required in training (see Fig.~\ref{fig:acc_ntraj}).
The simplicity of the resulting descriptors also allows features to be included at further distances from each particle,
allowing for the possibility of capturing structural correlations beyond each particle's immediate proximity.
Furthermore, as opposed to the traditional BP descriptors, features based on gaps and contacts do not contain any extra parameters in their definitions,
alleviating the need to fine tune the descriptor hyperparameters for each system of interest.

In the case of two-dimensional jammed packings of soft particles undergoing quasi-static shear, 
we find that $\delmin$ correlates strongly with a particle's susceptibility to rearrangements, 
providing an indirect measure of each particle's mobility.
We also find that the nearest neighbor environment -- a particle's species and the number of contacts with its neighbors --
contains most of the local structural information captured by softness in two dimensions.

For the physical problem of interest in this papers, we evaluate $\delmin$ for critical vibrational modes whose frequency vanishes at stress drops during athermal quasistatic shear.
The success of softness trained on $\delmin$ for predicting plastic events in this context suggests that local variations in the response to a rearrangement are closely related to particle mobility. 
This suggests that $\delmin$ could be applied to a variety of related physical systems as a means of providing insight into particle dynamics.
For example, $\delmin$ could be evaluated for any low-frequency vibrational modes -- not just the critical mode associated with an instability -- as a highly efficient way of extracting soft spots~\cite{Manning2011}. 
It could even be evaluated for the relative displacements between different configurations (for example, to the difference between two configurations separated by a strain step) to provide a potentially more useful measure of mobility than $\dmin$ in systems with inhomogeneous loads. It could also be evaluated to study the response of configurations with force dipoles applied in numerical simulation or even experiments.
Furthermore, one could use softness trained on $\delmin$ to predict rearrangements in athermal systems experiencing other types of loading such as uniform compression or expansion, or in thermal systems that are either quiescent or under load.

Although softness is highly useful as a structural predictor of mobility, there are situations where such a predictor is not needed and it may suffice to characterize the mobility. 
Elastoplasticity models are based on the premise that the coupling between mobility and elasticity is key to understanding the deformation and flow of disordered solids. Characterization of the interplay between $\delmin$ and elasticity and how this varies from system to system could lead to new elastoplasticity models based on relevant microscopic information.

In this study, we used the persistence analysis as a systematic means of identifying a set of local structural variables relevant to dynamics in jammed packings.
 The analysis can readily be extended to particles with more complicated sets of interactions. 
While it is always possible to form an unweighted Delaunay triangulation given just particle positions,
a cell complex (i.e., a generalization of a triangulation, see Ref.~\onlinecite{Edelsbrunner2010}) which corresponds more closely to the actual constraints or interactions in the system may provide cleaner results.
For example, one could imagine developing a generalization of the alpha-shape filtration, and associated triangulation, for non-spherical particles.
In lieu of a rigorous mathematical formulation, it would also be possible to pixelate the underlying space into a cubical complex and then evaluate the total potential energy on the vertices between pixels (or voxels)~\cite{Robins2011, Delgado-Friedrichs2015}.
Once could then perform a filtration of the potential energy function on this cell complex.
If the particles have well-defined boundaries, a Euclidean distance transform on the cubical complex could also suffice.
In all cases, once an appropriate filtration is chosen, the persistence algorithm will provide a complete characterization of the topological structure.

We have shown that $\delmin$ is a better dynamical quantity than $\dmin$ for determining softness from linear regression. For classification, $\delmin$ and $\dmin$ are equally effective. However, we note that for classification one could use local minima and maxima in $\delmin$ -- or even $\dmin$ directly -- 
as natural classes of non-rearrangers and rearrangers, respectively, instead of placing stringent thresholds on $\dmin$. This uses data more efficiently so that fewer snapshots are needed, although not as efficiently as linear regression.

Our result that gaps and contacts provide a concise and predictive characterization of local structure dovetails nicely with our understanding of jammed systems, where the contact number is a key quantity. 
Here we find that for predicting mobility, it is not only contacts that are important but also gaps in the Voronoi cell.
It would be interesting to determine whether gaps are themselves important
or are a signature of some other underlying aspect of local structure that is more closely related to the contacts.

While we find that we are able to achieve high percentiles of softness for the particles which experience the most motion,
we still do not achieve a perfect $100\%$ accuracy.
The fact that we observe $R^2$ values of only about $20\%$ provides 
a strong indication that our method still does not capture all of the relevant structural information.
The fact that we still achieve high accuracy for predicting the sources of rearrangements implies that these particles tend to be outliers in the distributions of local structures.

One important aspect of the local structure we neglect is the contact stresses between particles.
We note that although it is not utilized, the persistence analysis should capture this information in principle.
In fact, we observe a slight correlation of $\delmin$ with $\alpha_b$ and $\alpha_d$ in Figs.~\ref{fig:pdiag_D2min_normed}(c-i) and (c-iii) 
for both triangles with no gaps and cycles with more than three edges that contain no interior gaps.
The farther a feature is from the vertical $\alpha_b=0$ line, the smaller its average value of $\delmin$ seems to be.
That is, particles with in environments with larger contact overlaps, i.e. larger stress, seem to be less mobile.
In addition, the externally applied shear strain provides a natural anisotropy to the system,
which has been shown in other studies to be important in fully capturing which particles are likely to rearrange~\cite{Schwartzman-Nowik2019, Patinet2016}. 
However, we do not include any orientational information in our descriptors and the persistence algorithm we have demonstrated does not take orientation into account.
It would be useful to develop a way to either include this information in the persistent homology framework 
or at least find a means to correlate it with the features found by the standard algorithm.
The addition of information about stresses and orientation into our analysis would help to provide an upper limit
on the value of local structural information for the prediction of plastic rearrangements.

\section*{Acknowledgements}

This research was supported by the US Department of Energy, Office of Basic Energy Sciences,
Division of Materials Sciences and Engineering under Award DE-FG02-05ER46199 (J.W.R), 
the Natural Sciences and Engineering Research Council of Canada under a Postgraduate Scholarship -- Doctoral award (S.A.R.), 
and the Simons Foundation for the collaboration
Cracking the Glass Problem via award 454945 (J.W.R, S.A.R, and A.J.L) and Investigator Award 327939 (A.J.L.).

\section*{Data Availability}

The data that support the findings of this study are available from the corresponding author upon reasonable request.

\appendix

\section{Persistence Diagrams and Decompositions}

In this section, we provide any additional details necessary for producing the persistence diagrams and decompositions in Figs.~\ref{fig:pdiag} and \ref{fig:pdiag_D2min_normed}.

\subsection{Choosing Representative Cycles}\label{sec:repcycles}

In order to decompose the persistence diagrams in Figs~\ref{fig:pdiag} and \ref{fig:pdiag_D2min_normed},
we sort cycles according to size (number of edges), edge types (gaps vs. overlaps), and particle species (small vs. large radii).
However, the persistence algorithm does not provide a unique representation of the cycles it detects.
Instead, it only indicates when classes of homologous cycles appear and disappear during the course of the filtration.
This means that for any hole that appears in the union of discs,
there may be multiple cycles of edges in the triangulation that encircle that hole, resulting in the same birth-death pair.
Consequently, it is necessary to choose a particular representation of each cycle in order to classify them according to type.
For this study, the exact choice we make does not affect the overall results,
so we choose a representation of each cycle in a way that arises naturally from the persistence algorithm.

To perform the persistence algorithm, one starts by constructing the boundary matrix $\partial$, 
an operator which maps simplices (vertices, edges, triangles, etc.) in a cell complex to their boundaries.
Each row and column in $\partial$ represents a simplex, sorted in order of their appearance during the filtration 
(for simplicity, we will refer to the simplex represented by row or column $i$ as simplex $i$).
The $j$th row of $\partial$ is defined such that the element in the $i$th row is 
one if simplex $i$ is a boundary of simplex $j$ and zero otherwise.
Performing the persistence algorithm in our case amounts to transforming $\partial$ to Smith normal form via column additions and subtractions modulo 2.
In the resulting reduced matrix $R$, each column has a different pivot, the maximal row index of the nonzero column entries.
If a column $j$ has nonzero entries and its pivot is row index $i$, 
this means a feature was born upon the introduction of the simplex $i$ and died with simplex $j$ 
(see Ref.~\onlinecite{Edelsbrunner2010} for more detailed explanation).

The nonzero elements of a column $j$ in $R$ are a linear combination of simplices which form a cycle.
Since each of these nonzero elements has index less than or equal to the pivot, 
each of the cycle's constituent simplices were present at the time that feature was born.
Therefore, this cycle forms a representation of the topological feature at the time of birth, a birth cycle.
Furthermore, the boundary of the simplex $j$ forms a unique representation of the feature right before it's death, its death cycle.
This means that column $j$ is homologous to the birth cycle we have described.
In other words, at the time right before death, both the birth and death cycles surround the same hole in the triangulation.

Therefore, given a feature that is born with simplex $i$ and dies with simplex $j$,
we choose the $j$th column of the reduced boundary matrix $R$ as a representative cycle.
It is a cycle that both describes a feature when it is born and is homologous to the cycle at the time of death.
Furthermore, it is easy to compute, as it can simply be read off $R$ when computing the persistence algorithm with no modification.

We acknowledge that various methods exist to identify representative cycles.
For example, one could calculate optimal cycles, choosing the smallest representation of each cycle.
However, this method can be cumbersome, 
requiring the implementation of integer programming techniques~\cite{Escolar2016}.
A basis of birth cycles can also be found efficiently by finding the matrix $V$ 
such that $R=\partial V$ and reading off the columns corresponding to simplices at which features are born.
In both cases, the resulting cycles are not always guaranteed to be homologous to a feature at the time of death (or even birth for optimal cycles).
Our method of simply reading off the columns of $R$ does not suffer from any of these drawbacks.

\subsection{Triangular Cycle Persistence Curves}\label{sec:tricurves}
In this section, we derive the analytic forms of the persistence curves for the four type of triangles shown in Figs~\ref{fig:pdiag}(c-i)-(c-iv).
First, we introduce a general formalism for calculating $\alpha$-values for simplices embedded in $d$-dimensional space.
Next, we derive the solutions for the birth and death of triangular cycles with a single gap as a function of the size of the gap.

\subsubsection{Calculating $\alpha$-values}

Suppose we have a simplex consisting of $n$ points in $d$-dimensions 
($n \leq d+1$) with positions $\vec{v}^i$ where $i=1,\ldots, n$ with each point assigned a weight $w_i$. 
In the context of soft interacting spheres, we can write each weight as 
\begin{align}
w_i &= R_i^2 + \alpha
\end{align}
where $R_i$ is the interaction radius of particle $i$ and $\alpha$ is a scale factor 
used to control the radii of the balls when performing the filtration of the weighted Delaunay triangulation.

Next, we define the weighted squared distance, or power, of a point $\vec{x}$ from $\vec{v}^i$ as
\begin{align}
\pi_i(\vec{x}) &= \norm{\vec{x}-\vec{v}^i}^2 - w_i.
\end{align}
Note that $\pi_i(\vec{x}) = 0$ is the equation of a sphere centered at $\vec{v}^i$ with radius $\sqrt{w_i}$. 
The point $\vec{x}$ is said to be orthogonal to $\vec{v}^i$ if the power between the two points is zero.
We define a power sphere of a set of points as the $d$-dimensional sphere centered at $\vec{x}$ with $\vec{x}$ orthogonal to each point.

During a filtration on a Delaunay triangulation, the value of $\alpha$ at which a simplex comes into existence ,
i.e., that at which its balls all come into contact, is equivalent to that of the power sphere of its vertices.
The position of the power sphere is then the point at which each ball comes into contact.
Therefore, our goal is to find a position $\vec{a}$ and a scale factor $\alpha$ which define a power sphere for our $n$ points.
In the case where all points are equally weighted, this problem is equivalent to calculating the radius and position of a circumscribing $d$-sphere. 
If $n=d+1$, the radius is unique, but if $n < d+1$, we will choose the unique sphere with minimum radius.

First, for each point, we write down its orthogonality condition,
\begin{align}
\pi_i(\vec{a}, \alpha) &= \norm{\vec{a}-\vec{v}^i}^2 - R_i^2 - \alpha = 0.
\end{align}
Expanding the square, we obtain
\begin{align}
\norm{\vec{a}}^2 - 2\vec{v}^i\cdot \vec{a} + \norm{\vec{v}^i}^2  = R_i^2 + \alpha.
\end{align}
We note that this is a nonlinear equation for $\vec{a}$. 
To linearize, we define
\begin{align}
q &= \alpha - \norm{\vec{a}}^2\label{eq:linearize}
\end{align}
giving us
\begin{align}
2\vec{v}^i\cdot \vec{a} + q  = \norm{\vec{v}^i}^2 - R_i^2\label{eq:sphere}
\end{align}
which is a set of $n$ linear equations of $d+1$ unknowns, $\vec{a}$ and $q$. 

In the general case, we need an additional $n-(d+1)$ constraints to determine a unique solution. 
If we choose the minimum radius circumsphere, then its center will be coplanar with our $n$ points. 
Thus, we write the center of the sphere as a parametric function of the points,
\begin{align}
\vec{a} &= \vec{v}^1 + \sum\limits_{i=2}^n s_i (\vec{v}^i-\vec{v}^1)\label{eq:plane}
\end{align}
where we have introduced an additional $n-1$ free parameters $s_i$, $i=2,\ldots, n$. 
We now have a total of $d+1+n-1 = d+n$ free parameters $(\vec{a}, \{s_i\}, q)$. 
We also have an additional $d$ equations giving us a total of $d+n$, which means we have enough information to solve for the circumscribing power sphere. 
We then solve for $\vec{a}$ and $q$ using the system of linear equations represented by Eqs.~\ref{eq:sphere} and \ref{eq:plane} and obtain the scale factor using Eq.~\ref{eq:linearize}.
We note that if $n=d+1$, as is the case for a triangle in two-dimensions or a tetrahedron in three-dimensions, then we can omit Eq.~\ref{eq:plane}.
The equations derived here apply in dimensions $d \geq 1$ with $n\geq 2$ and can be used to construct a filtration on a weighted Delaunay triangulation.
For details of how to use $\alpha$-values to construct this filtration, see Refs.~\onlinecite{Edelsbrunner1992} and~\onlinecite{Edelsbrunner1992a}.

\subsubsection{Triangular Cycles ($d=2, n=3$)}

\begin{figure}[h!]
\centering
\includegraphics[width=0.75\linewidth]{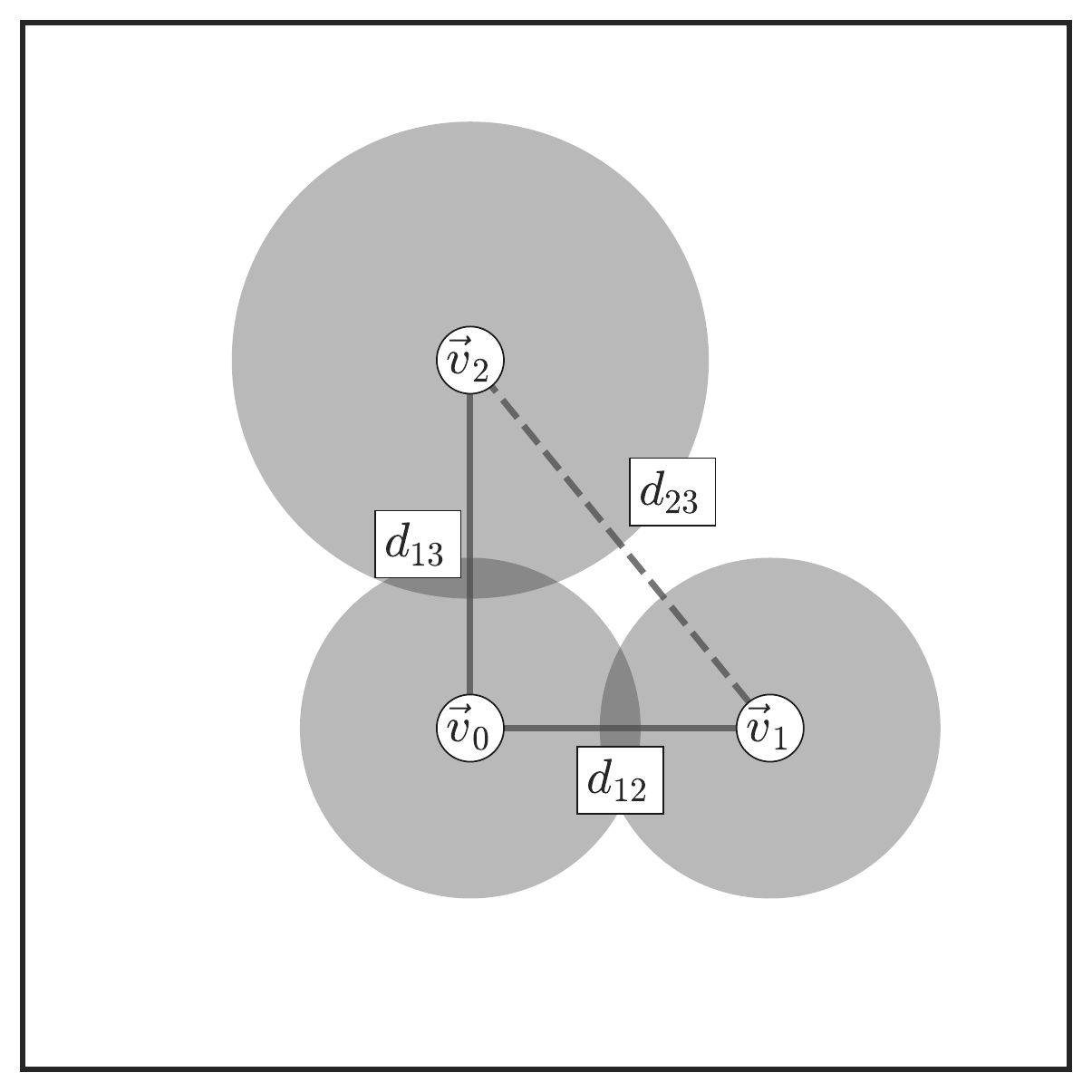}
\caption{Schematic of triangular cycle used to derive $\alpha_b$ and $\alpha_d$ as a function of gap size.}
\label{fig:triangle}
\end{figure}

Next, we use the formulation above to derive the values of $\alpha$ at which a cycle composed of three edges will be born $\alpha_b$ and die $\alpha_d$ during a filtration.
Suppose we have  $n=3$ particles in dimension $d=2$ with radii $R_1$, $R_2$ and $R_3$ and positions $\vec{v}^1$, $\vec{v}^2$ and $\vec{v}^3$, receptively.
Since only the positions of the particles relative to one another matter, without loss of generality, we write the particle positions as
\begin{equation}
\begin{split}
\vec{v}^1 &= (0,0)\\
\vec{v}^2 &= (r_{12}, 0)\\
\vec{v}^3 &= (r_{13}\cos\theta, r_{13}\sin\theta)
\end{split}
\end{equation}
where $r_{ij}$ is the Euclidian distance between particles $i$ and $j$ 
and $\theta$ is the angle of the triangle at the corner defined by particle $1$ such that
\begin{equation}
\begin{split}
\cos\theta &= \frac{r_{12}^2 + r_{13}^2 - r_{23}^2}{2r_{12}r_{13}}\\
\sin\theta &= \sqrt{1-\qty(\frac{r_{12}^2 + r_{13}^2 - r_{23}^2}{2r_{12}r_{13}})^2}.
\end{split}
\end{equation}
We wish to calculate $\alpha_b$ and $\alpha_d$ as a function of the triangle shape as one of the sides opens up into a gap.
Initially, when all three particles are in contact, we assume each pair of particles $i$ and $j$ overlaps by an amount $\delta_{ij} > 0$,
which can depend on particle species (see the next section for more details).
We place the gap between particles $2$ and $3$, parameterized by a parameter $\varepsilon$ such that at a minimum value of $\varepsilon=0$ all particles overlap by $\delta_{ij}$. 
All together, we parameterize the pairwise distances between particles as
\begin{equation}
\begin{split}
r_{12} &= R_1 + R_2 - \delta_{12}\\
r_{13} &= R_1 + R_3 - \delta_{13}\\
r_{23} &= R_2 + R_3 - \delta_{23} + \varepsilon.
\end{split}
\end{equation}
Fig.~\ref{fig:triangle} depicts a schematic of such a triangle.

From here, we calculate $\alpha_b$ and $\alpha_d$ as a function of the gap parameter $\varepsilon$.
First, to derive the birth of the triangle, we calculate $\alpha$ for each edge in the triangle ($n=2$).
We use Eqs.~\ref{eq:sphere} and \ref{eq:plane}, along with Eq.~\ref{eq:linearize} to solve for $\alpha_{ij}$ for an edge between particles $i$ and $j$, resulting in
\begin{align}
\alpha_{ij} &=  -R_i^2 + \frac{1}{4r_{ij}^2} \qty[r_{ij}^2 + R_i^2 - R_j^2]^2.
\end{align}
This equation can be shown to be symmetric in $i$ and $j$.
The triangle will be born when all three possible pairs of particles start to overlap.
Consequently, we take the maximum value $\alpha_{ij}$ out of all three pairs:
\begin{align}
\alpha_b &=  \max \qty(\alpha_{12}, \alpha_{13}, \alpha_{23}).
\end{align}

Next, we derive the the value of $\alpha$ at which the cycle defined by the triangle dies during the filtration.
For a triangle in two-dimensions, $n=d+1$ and Eqs.~\ref{eq:sphere} and \ref{eq:linearize} are sufficient to solve for $\alpha$,
with the result
\begin{equation}
\begin{split}
\alpha_d(\varepsilon) =& -R_1^2 + \frac{1}{4r_{12}^2} \qty[r_{12}^2 + R_1^2 - R_2^2]^2 \\
&+ \left(\frac{1}{2r_{13} \sin\theta}\qty[r_{13}^2 + R_1^2-R_3^2]\right. \\
&\left.-\frac{1}{2r_{12} \tan\theta}\qty[r_{12}^2 + R_1^2-R_2^2]\right)^2.
\end{split}
\end{equation}
The persistence curve is defined parametrically as a function of $\varepsilon$ by following the path of the point $(\alpha_b(\varepsilon), \alpha_d(\varepsilon))$
for a particular combination of particle sizes.
The gap parameter $\varepsilon$ starts at a value of $\varepsilon_{\min}=0$.
As $\varepsilon$ increases and a triangle is stretched out, eventually, the difference between the birth and death $\alpha$-values of the triangle will approach one another.
If there is a point at which they coincide, then the maximum valid value of the gap parameter, $\varepsilon_{\max}$ will be the solution to the equation
\begin{align}
\alpha_b(\varepsilon_{\max}) = \alpha_d(\varepsilon_{\max}).
\end{align}
If a solution to this equation does not exist, then $\varepsilon_{\max}$ will be the point at which they are closest together defined by
\begin{align}
\varepsilon_{\max} = \argmin_\varepsilon \qty[\alpha_d(\varepsilon) - \alpha_b(\varepsilon)].
\end{align}

This derivation can easily be extended to higher-dimensional simplices such as tetrahedra, and also to higher embedding dimensions.
In addition, one could choose different values of the initial overlap between particles or could explore the area swept out in the persistence diagrams
by adding more than one gap to a simplex.

\subsubsection{Estimating Contact Overlap}

In the previous section, we introduced the contact overlap $\delta_{ij}$ for a pair of particles $i$ and $j$.
In general, the average value of this overlap will depend on the details of the interaction between the particles.
Here we choose to estimate this parameter by relating it to the contact forces between particles.
We start by calculating the force between particles $i$ and $j$ by taking the derivative of the potential in Eq.~\ref{eq:potential}
with respect to the interaction distance (we have dropped the Heaviside function),
\begin{align}
f(r_{ij}) = -\frac{\epsilon}{(R_i+R_j)}\qty(1 - \frac{r_{ij}}{R_i+R_j} )^\frac{3}{2}.
\end{align}
Next, plugging in the relation between particle distance and contact overlap, $r_{ij} = R_i + R_j - \delta_{ij}$, and solving for $\delta_{ij}$, we obtain
\begin{align}
\delta_{ij} = \qty(-\frac{f}{\epsilon})^\frac{2}{3} (R_i+R_j)^\frac{5}{3}.
\end{align}
This relationship holds for any combination of particle sizes with only one parameter that must be specific $f/\epsilon$.
We estimate this parameter from our configurations by calculating the average force between particles that are in contact.
We measure this quantity to be approximately $f/\epsilon\approx-0.0093$.

\section{Softness Calculation Details}

\subsection{Structural Descriptors}\label{sec:descriptors}

\subsubsection{Behler-Parrinello Descriptors}\label{sec:BP}

The Behler-Parinellow structural descriptors~\cite{Behler2007} provide a parameterization of each particle's local structure.
For a particle $i$ we define the radial descriptors as
\begin{align}
G_Y^X(i; \mu) = & \sum_j e^{-(r_{ij}-\mu)^2/L^2}
\end{align}
where $j$ sums over all particles, $r_{ij}$ is the distance between particles $i$ and $j$, $X$ and $Y$ indicate particle species, and $\mu$ and $L$ are constants.
For a pair of particles of species $X$ and $Y$ with combined radii $\sigma_{\mathrm{tot}} = R_X+R_Y$, we use values of $\mu$ ranging from $0.8\sigma_{\mathrm{tot}}$ to $2.0\sigma_{\mathrm{tot}}$ in steps of $0.05\sigma_{\mathrm{tot}}$ with $L=0.05\sigma_{\mathrm{tot}}$.
The angular descriptors are defined as
\begin{equation}
\begin{split}
\Psi_{YZ}^X(i; \xi, \lambda, \zeta)=& \sum_{jk} e^{-(r^2_{ij} + r^2_{ik} + r^2_{jk})/\xi^2}\\
&\times\qty(1+\lambda \cos \theta_{ijk})^\zeta
\end{split}
\end{equation}
where $\xi$, $\lambda$ and $\zeta$ are constants and $\theta_{ijk}$ is the angle at the corner $i$ of the triangle defined by particles $i$, $j$ and $k$.
We use the same set of values for the four parameters as Ref.~\onlinecite{Cubuk2015}.
Combining all parameters, we construct a vector $\vec{x}$ of descriptors in a similar manner to the gaps and contacts explained in the main text.

To calculate an representative average distance of each descriptor from a central particle $i$,
we treat each descriptor as a type of integration kernel, averaging the distance $r_{ij}$ over all $N$ particles under consideration.
The average distance of the radial descriptors for a particular value of $\mu$ is given by
\begin{align}
\expval{r}_{G(\mu)} =& \frac{1}{N}\sum_{i, j} r_{ij}e^{-(r_{ij}-\mu)^2/L^2}
\end{align}
while the average distance of an angular descriptor for $\xi$, $\lambda$ and $\zeta$ is similarly
\begin{equation}
\begin{split}
\expval{r}_{\Psi(\xi, \lambda, \zeta)} =& \frac{1}{N}\sum_{ijk}\frac{1}{2}(r_{ij} + r_{ik}) e^{-(r^2_{ij} + r^2_{ik} + r^2_{jk})/\xi^2}\\
&\times\qty(1+\lambda \cos \theta_{ijk})^\zeta
\end{split}
\end{equation}
where we have averaged over $r_{ij}$ and $r_{ik}$ to maintain symmetry.

\subsubsection{Gaps and Contacts}

The definition of the gap and contact descriptors are given in the main text.
To calculate the average distance of each descriptor, we first calculate the Euclidean distance of each edge from each particle in the Delaunay triangulation of our configurations.
The position of an edge is taken as the midpoint between its defining pair of particles.
We then average this distance separately for gaps and contacts at each  discrete triangulation distance $d_{i, (j,k)}$ as defined in Eq.~\ref{eq:dist}.

\subsection{Classification}

\subsubsection{Training Set Construction}

To construct our training set for classification,
we first calculate  $\dmin$ for each configuration and sort the particles in increasing order.
Next, we convert this ordering 
to the quantile of each particle $i$ within that configuration, denoted $q_i$, which ranges from 0 to 1.
Examples of soft particles are then chosen as particles where $q_i$ is greater than  $q_{\text{r}}$, 
the upper quantile threshold.

To select hard particles, we use a slightly different approach.
For each particle in a particular rearrangement configuration, we record the maximum value of $\dmin$ experienced by that particle within a window of 10 future rearrangements (including the current rearrangement).
Next, we again convert this quantity to a quantile representation $q_i'$ and choose all particles with $q_i'$ less than $q_{\text{nr}}$ as examples of hard particles.
In this work, we always choose $q_{\text{r}} = q_{\text{nr}}$.

\subsubsection{Model}

To perform classification, we utilize a support vector machine (SVM).
In this framework, each particle $i$ has a label $y_i$ where  $y_i = -1$ for soft particles and $y_i = 1$ for hard particles,
along with a vector of features $\vec{x}_i$, which may be BP descriptors or our descriptors based on gaps and contacts.
We define $N$ to be the number of particles used in training.
Training the classifier than equates to solving the following optimization problem for the vector of weights $\vec{w}$, intercept $b$, and slack variables $\zeta_i$:
\begin{equation}
\begin{gathered}
min_{\vec{w}, b, \vec{\zeta}} \frac{1}{2} \norm{\vec{w}}_2^2 + C \sum_{i=1}^N \zeta_i\\
\text{subject to }  y_i(\vec{w}\cdot \vec{x}^i + b) \geq 1 - \zeta_i,\\
\zeta_i \geq 0, \quad i=1\ldots N
\end{gathered}
\end{equation}
The hyperparameter $C$ controls regularization.
This formulation equates to finding a hyperplane with normal vector $\vec{w}$ which best separates the two classes of particles in the space of features.
We then calculate the softness $S_i$ for each particle as a weighted sum of features,
\begin{align}
S_i = \vec{w}\cdot\vec{x}^i.
\end{align}

\subsection{Regression}

The formulation for regression we use is standard ridge regression.
For each particle in the training set, we have an independent value $y_i$ given by $\delmin(i)$ along with a vector of features $\vec{x}_i$, which may be BP descriptors or our descriptors based on gaps and contacts.
Defining $N$ as the number of particles used in training,
we perform the following optimization problem for the vector of weights $\vec{w}$ and intercept $b$:
\begin{align}
\min_{\vec{w}, b} \sum_{i=1}^N (\vec{w}\cdot\vec{x}^i + b - y_i)^2 + \alpha \norm{\vec{w}}^2
\end{align}
The hyperparameter $\alpha$ controls regularization.
W then calculate the softness $S_i$ for each particle as a weighted sum of features,
\begin{align}
S_i = \vec{w}\cdot\vec{x}^i.
\end{align}

\subsection{Machine Learning Protocol}\label{sec:protocol}

We perform most of our machine learning tasks using the scikit-learn Python package~\cite{scikit-learn}.
Our learning protocol consists of the following three steps steps:
\begin{enumerate}
\item Rescale all structural descriptors to zero mean and unit variance.
\item Determine optimal hyperparameter values using cross-validation.
\item Fit the model and use cross-validation or bootstrapping to calculate the mean and standard deviation of the test accuracy.
\end{enumerate}
We elaborate on these steps in the following sections.

\begin{figure}[h]
\centering
\includegraphics[width=0.95\linewidth]{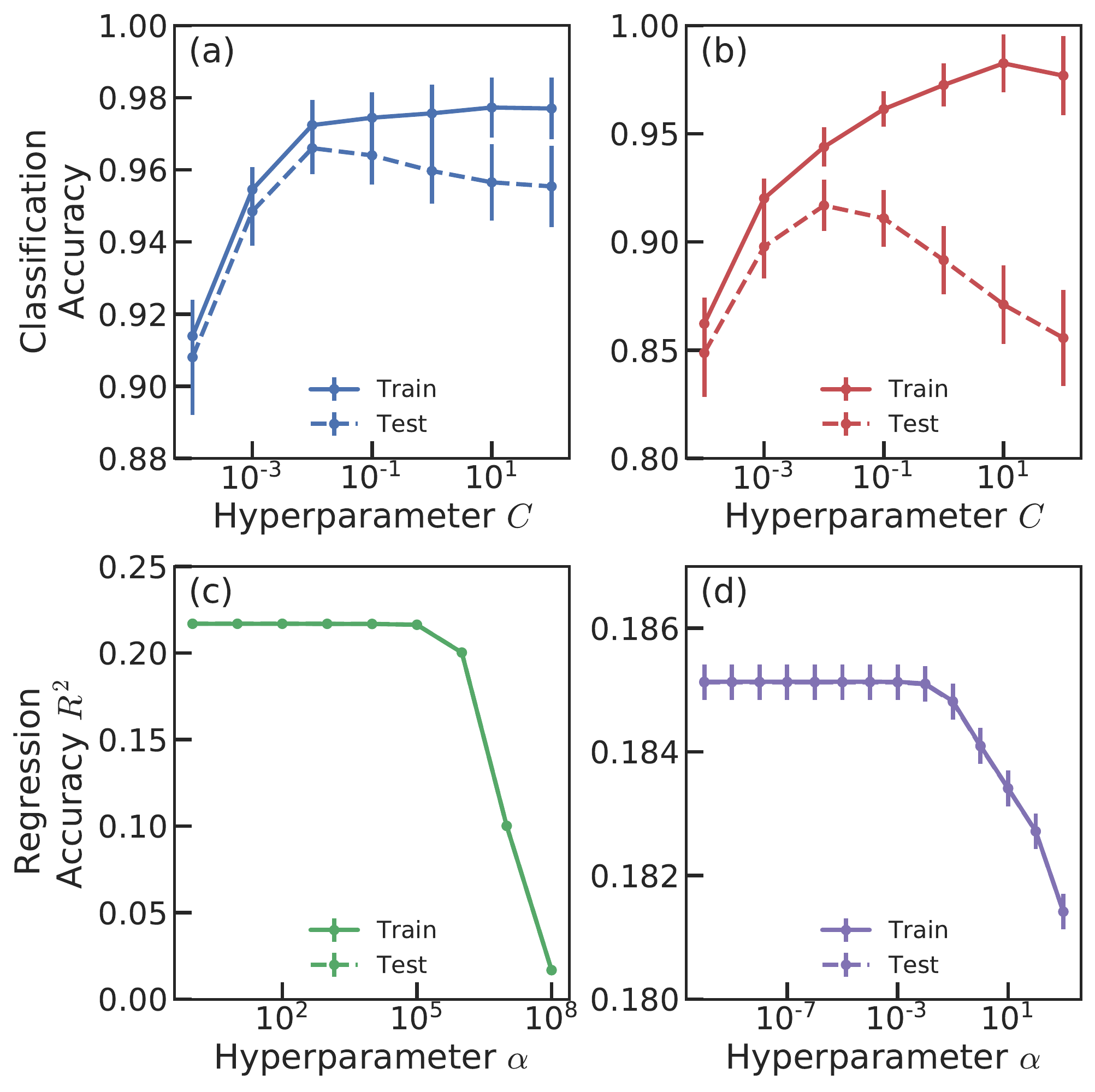}
\caption{Search for optimal hyperpameters the four different models.
In all cases, we show the training and test accuracies (solid and dashed lines, respectively) relevant to the model type.
Classification accuracy is shown as a function of the regularization paramter $C$ for (a) descriptors based on gaps and contacts and (b) BP descriptors.
Similarly, regression accuracy is shown as a function of the regularization parameter $\alpha$ for (a) gaps and and contacts and (b) BP descriptors.}
\label{fig:hyper}
\end{figure}

\subsubsection{Descriptor Rescaling}

Before training either of our models, we independently standardize each structural descriptor.
To standardize a descriptor, we subtract its observed median value and divide the descriptor by the interquartile range (IQR), the difference between the 25th and 75th percentiles.
We determine the median and IQR only considering data in the training set within a particular cross-validation or bootstrapping set, so as to avoid over-fitting.
We use the median and IQR to standardize so as to reduce the potential influence of outliers.

\subsubsection{Cross-validation}

To avoid overfitting, we perform repeated two-fold cross-validation in which we randomly sort our trajectories into two sets of configurations, a training set and a test set, both of equal size.
We then fit our model using the training set data and evaluate accuracy on the test set.
We repeat this process $32$ times and measure the mean and variance of the resulting accuracies.

\subsubsection{Hyperparameter Search}

For each of the two model types, we optimize one hyperparamter: $C$ for classification or $\alpha$ for regression.
To determine the best value for a hyperparamter, we scan through a range of values spaced on a log-scale and calculate a cross-validated train and test accuracies at each value.
We then choose the parameter which results in the largest test accuracy. 
Fig.~\ref{fig:hyper} shows the range of values scanned for the primary models reported in the main text.
For regression we chose $\alpha = 10^3$ when using gaps and contacts and $\alpha = 10^{-5}$ when using BP descriptors.
For classification, we find that the choice of $q$ does not have a significant affect on this optimal value.
In Figs.~\ref{fig:hyper}(a) and (b), we have shown results for $q=10^{-4.5}$, corresponding to one particle of each class per configuration.
For this model type, we chose $C = 10^{-2}$ for both types of descriptors.
In principle, the cross-validation could also be performed for the discrete radius of the neighborhood used to calculate $\dmin$ and separately the discrete neighborhood radius used to rescale $\dmin$ to calculate $\delmin$.
In addition, there are many choices for the hyperparameters that are used to define the BP descriptors.

\subsubsection{Computing Model Accuracy}

After performing our hyperparameter search, we take note of whether there is a significant difference between the training and test accuracies. 
If there is a significant difference for a particular model, we use cross-validation when we later evaluate the success of that model, reporting the relevant accuracy and $\CDF$ values as measured on the test set.
If there is no significant difference, then we perform bootstrapping to evaluate success, as it requires significantly less computational power.
When performing bootstrapping, we randomly sample configurations with replacement.
We then fit the model to this resampled data set and evaluate accuracy and $\CDF$ on the same dataset.
In both cases, we always resample our data set 32 times.
In Figs.~\ref{fig:hyper}(a) and (b), we see that for classification, the test accuracy is generally less than the train accuracy.
This means that we always use cross-validation to evaluate success for classification-based schemes.
In contrast, we see in Figs.~\ref{fig:hyper}(c) and (d), the two accuracies do not differ significantly for regression.
We therefore use bootstrapping to evaulate success when performing regression.


%

\end{document}